%% file: main.tex
\documentclass[sigconf,screen,nonacm]{acmart}
\setcopyright{none}
\renewcommand\footnotetextcopyrightpermission[1]{} 

\input{_global_settings}
\begin{document}

\title{\kjc: Automated Vulnerable Environment Generation for Linux Kernel Vulnerabilities}
\author{Bonan Ruan}
\affiliation{%
  \institution{National University of Singapore}
  \city{}
  \country{}
}

\author{Jiahao Liu}
\affiliation{%
  \institution{National University of Singapore}
  \city{}
  \country{}
}

\author{Chuqi Zhang}
\affiliation{%
  \institution{National University of Singapore}
  \city{}
  \country{}
}

\author{Zhenkai Liang}
\affiliation{%
  \institution{National University of Singapore}
  \city{}
  \country{}
}

\input{sections/00-abstract}

\ccsdesc[500]{Security and privacy~Vulnerability management}
\ccsdesc[500]{Security and privacy~Software security engineering}

\keywords{Vulnerable Environment; Reproduction; Linux Kernel}

\maketitle

\input{sections/01-introduction}
\input{sections/02-background}
\input{sections/03-motivations}
\input{sections/04-approach}
\input{sections/05-implementation}
\input{sections/06-evaluation}
\input{sections/07-discussion}
\input{sections/08-related}
\input{sections/09-conclusion}

\input{main.bbl}


\input{sections/99-appendix}

\end{document}

%% file: _global_settings.tex
\let\checkmark\undefined
\usepackage{dingbat}
\usepackage{listings}
\usepackage{color}
\usepackage{xcolor}
\usepackage{multirow}

\usepackage[linesnumbered,ruled,boxed,lined]{algorithm2e}
\SetAlgoVlined 
\SetAlgoLined 
\SetKwInput{KwInput}{Input}  
\SetKwInput{KwOutput}{Output}  
\SetKwProg{Proc}{Procedure}{}{}

\DontPrintSemicolon

\usepackage{algpseudocode}
\usepackage{url}
\usepackage{float}
\usepackage{hyperref}
\usepackage{caption}

\algtext*{EndProcedure}

\definecolor{mypurple}{HTML}{67379A}
\definecolor{myyellow}{HTML}{FAD978}
\definecolor{myblue}{HTML}{93AAD8}
\definecolor{myred}{HTML}{EB3323}
\definecolor{mygreen}{HTML}{4EAD5B}
\definecolor{greenbg}{rgb}{0.9, 1, 0.9}

\lstdefinestyle{mystyle}{
    backgroundcolor=\color{white},   
    basicstyle=\ttfamily\footnotesize,
    breakatwhitespace=false,         
    breaklines=true,                 
    captionpos=b,                    
    keepspaces=true,                 
    numbers=left,                    
    numbersep=5pt,                  
    showspaces=false,                
    showstringspaces=false,
    showtabs=false,                  
    tabsize=2,
    numberstyle=\tiny\color{black},
    framexleftmargin=5mm, frame=single, framerule=0pt,
    xleftmargin=5mm,
    xrightmargin=3mm,
    framesep=5mm, 
    rulecolor=\color{black},
    postbreak=\mbox{\textcolor{red}{$\hookrightarrow$}\space},
}
\usepackage{amsthm}
\theoremstyle{definition}
 
\theoremstyle{plain}

\newcommand{\kjc}{\textsc{KernJC}\xspace}

\usepackage{pifont}
\newcommand{\xmark}{\ding{55}} 

\newcommand{\dc}{$D$\xspace}
\newcommand{\hc}{$H$\xspace}

\newcommand{\graph}{$G$\xspace}
\newcommand{\ddc}{$DDC$\xspace}
\newcommand{\dpc}{$DPC$\xspace}
\newcommand{\dcc}{$DCC$\xspace}
\newcommand{\hrc}{$HRC$\xspace}
\newcommand{\hsc}{$HSC$\xspace}
\newcommand{\hdc}{$HDC$\xspace}

\makeatletter
\newcommand{\removelatexerror}{\let\@latex@error\@gobble}
\makeatother

%% file: sections/00-abstract.tex
\begin{abstract}
Linux kernel vulnerability reproduction is a critical task in system security.
To reproduce a kernel vulnerability, the vulnerable environment and the Proof of Concept (PoC) program are needed.
Most existing research focuses on the generation of PoC, while the construction of environment is overlooked.
However, establishing an effective vulnerable environment to trigger a vulnerability is challenging. 
Firstly, it is hard to guarantee that the selected kernel version for reproduction is vulnerable, as the vulnerability version claims in online databases can occasionally be spurious.
Secondly, many vulnerabilities can not be reproduced in kernels built with default configurations. Intricate non-default kernel configurations must be set to include and trigger a kernel vulnerability, but less information is available on how to recognize these configurations. 

To solve these challenges, we propose a patch-based approach to identify real vulnerable kernel versions and a graph-based approach to identify necessary configs for activating a specific vulnerability.
We implement these approaches in a tool, \kjc, automating the generation of vulnerable environments for kernel vulnerabilities. 
To evaluate the efficacy of \kjc, we build a dataset containing 66 representative real-world vulnerabilities with PoCs from kernel vulnerability research in the past five years. 
The evaluation shows that \kjc builds vulnerable environments for all these vulnerabilities, 48.5\% of which require non-default configs, and 4 have incorrect version claims in the National Vulnerability Database (NVD).
Furthermore, we conduct large-scale spurious version detection on kernel vulnerabilities and identify 128 vulnerabilities which have spurious version claims in NVD.
To foster future research, we release \kjc with the dataset in the community.
\end{abstract}

%% file: sections/01-introduction.tex
\section{Introduction}
\label{sec:intro}
The Linux system has become one of the cornerstones of our modern computing infrastructure, supporting a wide range of devices and services, such as cloud servers and containers, Android devices, and IoT nodes. Breaches of Linux systems can precipitate catastrophic consequences. 

\begin{figure}[t]
  \centering
  \includegraphics[width=0.92\linewidth]{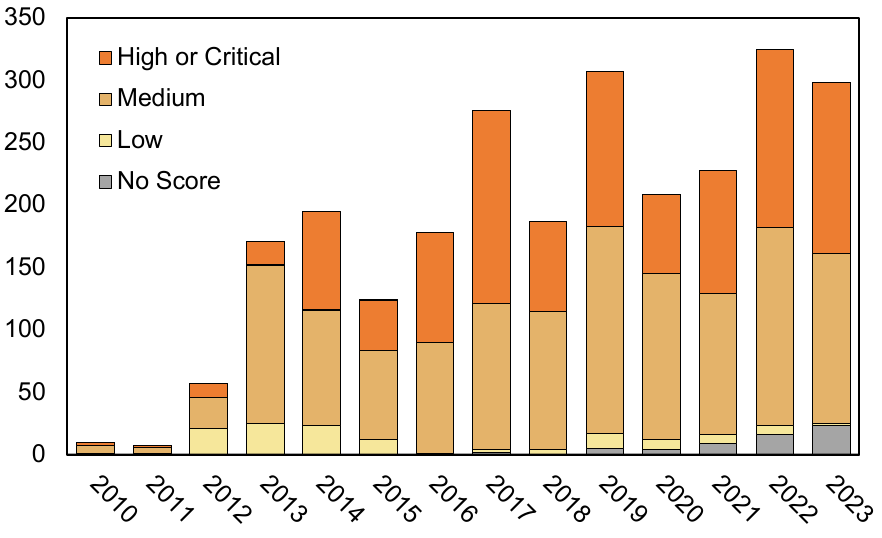}
  \caption{Numbers of Kernel Vulnerabilities since 2010}
  \label{fig:kernel_vuln_num}
\end{figure}

A primary attack surface within the Linux ecosystem is the Linux kernel, or the \textit{kernel}. 
It is the ultimate line of defense and the gatekeeper of Linux system security, where the exploitation of kernel vulnerabilities could result in various severe impacts, such as privilege escalation and denial of services on traditional servers, rooted Android devices~\cite{xu2015collision}, and container escaping in cloud-native environments~\cite{yang2021security}.
As a substantial and complex project, the Linux kernel comprises over 28 million lines of code~\cite{phoronixLinux512}. 
Kernel vulnerabilities can have far-reaching impacts, while their emergence seems endless. 
As shown in \autoref{fig:kernel_vuln_num}, the annual number of reported vulnerabilities in the upstream kernel has been trending upwards, with a notable increase in high or critical severity vulnerabilities~\cite{linuxkernelcvesLinuxKernel}, which accounts for more than 40\% in 2023. 
This trend underscores an escalating risk profile, necessitating sustained and focused security measures to safeguard systems against exploitation.

Kernel vulnerability reproduction and replay are often an essential part of assessing the vulnerability's severity and impact, designing solutions and mitigation techniques, and evaluating the effectiveness of solutions. 
Also, the reproducibility of vulnerabilities emerges as a crucial factor in their prioritization.
Prior studies indicate that vulnerabilities lacking reproducibility are often overlooked~\cite{mu2018understanding, votipka2018hackers}, thereby leaving systems exposed to potential threats. 
Furthermore, by reproducing kernel vulnerabilities, analysts can comprehend the attack behaviors at runtime to update the intrusion detection systems, which plays a pivotal role in detecting and preventing future vulnerability exploitation attacks.

To reproduce a kernel vulnerability, there are two crucial elements: the vulnerable environment, and the Proof of Concept (PoC) program. 
The role of the vulnerable environment is to guarantee the existence and accessibility of the vulnerability in question, thereby establishing a suitable setting for analysis.
On the other hand, the PoC is specialized for triggering this vulnerability.
By executing the PoC in the environment, security analysts complete the reproduction and then gain further intelligence for the vulnerability.
The majority of existing research tends to concentrate primarily on the intricacies and development of PoC~\cite{you2017semfuzz,chen2020systematic,chen2019slake,liu2022release,wang2023alphaexp,chen2020koobe,wu2019kepler,wu2018fuze,jiang2023aem}. However, constructing the vulnerable environment is also an important but overlooked direction. Several solutions~\cite{chen2020systematic,chen2019slake,liu2022release,wang2023alphaexp} are concentrated on the static and dynamic identification of critical kernel objects to bypass specific mitigation mechanisms and facilitate kernel vulnerability exploitation. You \textit{et al.}~\cite{you2017semfuzz} focuses on the automated generation of PoC for kernel vulnerabilities. ~\cite{chen2020koobe,jiang2023aem,liu2022erace,wu2019kepler,wu2018fuze} concentrate on the automated generation and migration of ExPloit (ExP) for certain kernel vulnerabilities. 

However, constructing an appropriate reproduction environment for a given Common Vulnerabilities and Exposures (CVE) ID often presents considerable challenges~\cite{chen2021robin, mu2018understanding}.
In particular, there are two primary challenges. First, it is challenging to guarantee that the selected kernel version for reproduction is vulnerable, in that the claimed vulnerable versions in online databases, such as the National Vulnerability Database (NVD)~\cite{nvd}, may occasionally be erroneous~\cite{bao2022v}, leading to tremendous efforts wasted on building and testing non-vulnerable kernels. For example, the NVD's claim regarding kernels up to v5.12 being vulnerable to CVE-2021-22555~\cite{nistCVE202122555} is inaccurate; kernels from v5.11.15 to v5.11.22, within the alleged range, have already been patched. The real latest vulnerable version is v5.11.14. Second, lots of vulnerabilities can not be reproduced in kernels built from default upstream configurations (referred to hereafter as \textit{configs}). it is time-consuming to explore the configs to identify and enable the necessary kernel configs responsible for the activation of a specific vulnerability, due to the complexity of the kernel configuration system.

Vulnerabilities often occur in specific subsystems or as a result of particular module features, which may not be active by default. Nevertheless, this important detail is frequently omitted in both CVE databases and external reports, complicating the process of activating kernel vulnerabilities. What's worse, the reliability of config files from successful reproduction attempts is also questionable, as evidenced by sources like~\cite{bsaucekernelExploitCVE202122555, githubkernelExploitCVE20220185} and our observations of vulnerability disclosure \cite{seclistsCVE202232250, openwallOsssecurityCVE202332233} in \autoref{sec:case-bpf-netfilter}. Similar to the situation in~\cite{chen2021robin}, it is not practical to enable all kernel configs as well, as some configs are mutually exclusive, and others may affect the execution of target functionality. Although there is an \verb|allyesconfig| option available in the Linux kernel build mechanism enabling as many configs as possible, our manual experiments and prior work~\cite{hao2023syzdescribe, you2017semfuzz} confirm that the kernel built with \verb|allyesconfig| is not bootable.

In this paper, we introduce \kjc\footnote{\kjc stands for Kernel JiaoChang, where JiaoChang, in ancient China, referred to a site dedicated to military training and competition.}, a novel tool that adopts a patch-based approach for pinpointing inaccuracies in version claims of online databases and accurately determining the actual vulnerable kernel version. Additionally, \kjc utilizes a graph-based approach to autonomously ascertain the specific kernel config set required for each vulnerability. \kjc also provides an intuitive command-line interface for managing kernel vulnerabilities and deploying PoCs.

Our implementation of \kjc, when applied to analyze kernel vulnerabilities, identifies 128 instances of false positive version claims in the NVD database. Furthermore, we test \kjc against 66 kernel vulnerabilities associated with PoCs in prominent research publications~\cite{chen2019slake, lee2021exprace, jiang2023aem, wu2019kepler, lee2023pspray, chen2020koobe, wang2023alphaexp, lin2022dirtycred, zeng2022playing, liu2022linkrid, abubakar2021shard, chen2020systematic, zeng2023retspill, wang2023pet, bhattacharyya2022midas, jeong2023segfuzz, jeong2019razzer, li2023hybrid, yoo2022kernel, yuanddrace, zhao2022statefuzz, pitigalaarachchi2023krover} from main security conferences over the past five years. The evaluation results reveal \kjc's efficacy in accurately establishing vulnerable environments and accomplishing the reproduction for all these vulnerabilities, with 48.5\% requiring non-default kernel configs. Notably, \kjc also uncovered 4 out of the 66 vulnerabilities that have incorrect version claims in the NVD database.

To the best of our knowledge, this is the first work focusing on the automated generation of the reproduction environment for Linux kernel vulnerabilities. 
In particular, our work makes the following contributions:

\begin{itemize}
 \item We develop techniques to correct and enhance the information available for kernel CVEs, fixing version errors and adding essential configs needed by reproduction. Especially, our research finds 128 cases of incorrect vulnerability version claims in the NVD database, emphasizing the need to improve vulnerability reporting accuracy. We implement our solution as an open-source tool, \kjc.
 
 \item We evaluated \kjc with representative real-world kernel vulnerabilities. The evaluation showcases \kjc's efficacy in constructing and automating the setup of vulnerable environments, where the target vulnerabilities are available and accessible from userland.
  
\item We have compiled and openly shared a dataset comprising 2,256 kernel vulnerabilities, 1,829 of which have verified vulnerable versions (whose version ranges are claimed in NVD database), 1,633 of which have identified kernel configs (exclusive of vulnerabilities that do not rely on any kernel config), and 66 of which are paired with functional PoCs. This dataset is currently the most extensive collection of kernel vulnerabilities and their PoCs known to us.

\end{itemize}

%% file: sections/02-background.tex
\begin{figure*}[t]
  \centering
  \includegraphics[width=\textwidth]{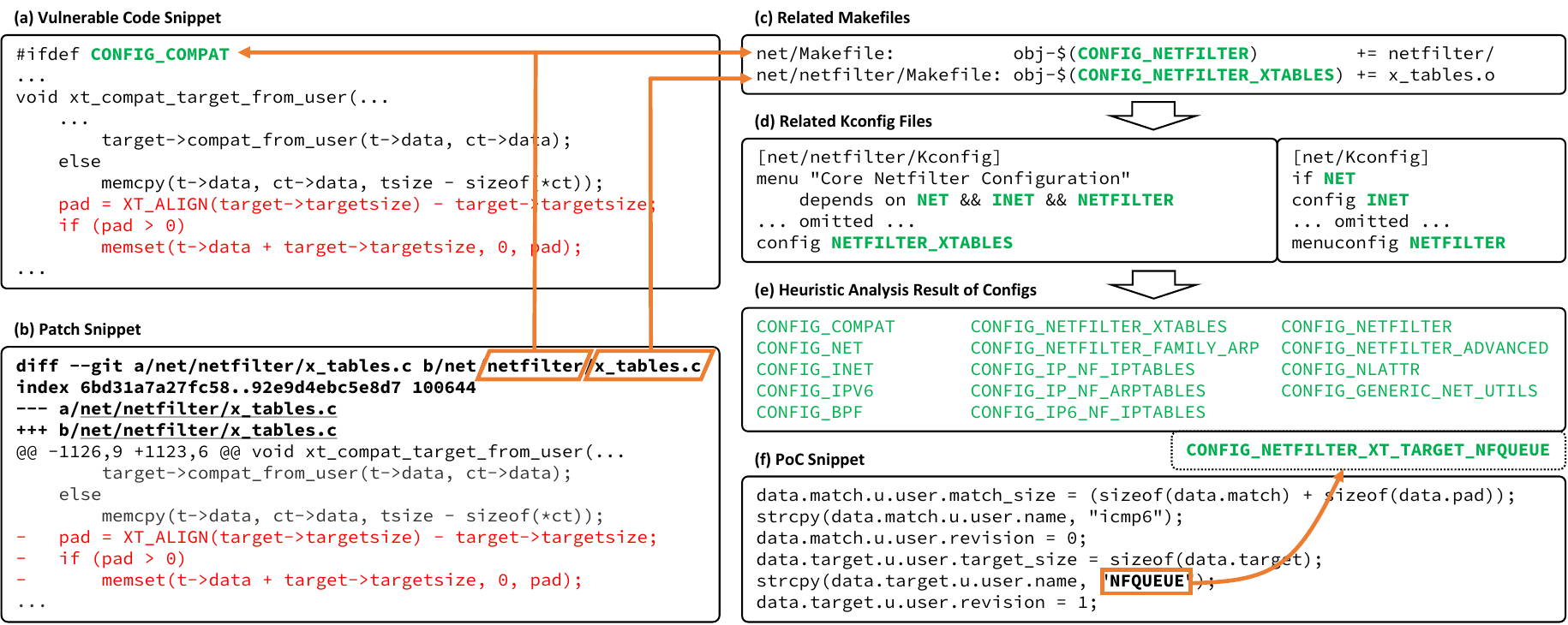}
  \caption{An Example Vulnerability from CVE-2021-22555. (a) shows one snippet from the source code of this vulnerability in the v5.11.14 kernel. (b) shows one snippet of the patch for the vulnerability. (c) shows the Makefiles related to the source code snippet from \lstinline{net/netfilter/x_tables.c}. (d) shows the Kconfig files related to \lstinline{net/netfilter/x_tables.c}. (e) shows the heuristic config analysis result based on the paths and contents of vulnerable files, patch, and the Kconfig\&Kbuild mechanisms. (f) shows one snippet from the public PoC~\cite{poccve202122555} for this vulnerability. The orange boxes and arrows illustrate the derivation of configs.}
  \label{fig:cve-2021-22555-motiv}
\end{figure*}

\section{Background and Motivation\label{sec:background}}

In this section, we first introduce the process of kernel vulnerability reproduction, emphasizing the core focus of our study --- constructing environments susceptible to kernel vulnerabilities.
Following this, we use \texttt{CVE-2021-22555} as a concrete example to illustrate the challenges encountered in creating such vulnerable environments.

\subsection{Kernel Vulnerability Reproduction}
\label{sec:lkvr}

Once a kernel vulnerability is disclosed, the standard reproduction process by analysts involves two key stages: constructing the corresponding vulnerable kernel environment and developing a workable Proof of Concept (PoC) program.

The goal of constructing vulnerable environments is to create an environment with a specific vulnerability accessible to users for interaction or testing.
This process requires the analyst to identify a kernel version known to be vulnerable, enable necessary configs, and compile the kernel image. 
Subsequently, the analyst has two options: either install the kernel on a physical machine and reboot to activate the vulnerable version or utilize virtualization tools like QEMU~\cite{bellard2005qemu} to set up a virtual machine tailored for the reproduction task.
In the virtualization scenario, integrating a corresponding root file system with the kernel image is crucial for a fully operational environment.
A prevalent method for constructing the root file system is using the script from the syzkaller project~\cite{githubSyzkallertoolscreateimageshMaster}, a well-known unsupervised coverage-guided kernel fuzzer~\cite{githubGitHubGooglesyzkaller}.

PoC development involves crafting a functional PoC using available vulnerability descriptions, patches, and technical reports.
More specifically, a PoC is a custom program created to attack the kernel-vulnerable environment, often leading to kernel crashes or inducing a certain kernel state (\textit{e.g.,} terminated), thereby demonstrating the reproducibility of a vulnerability.

Successfully reproducing a kernel vulnerability is the cornerstone of assessing its severity~\cite{mu2018understanding}.
Recent efforts have primarily concentrated on the development of PoC exploits to trigger these vulnerabilities~\cite{you2017semfuzz,chen2020koobe,jiang2023aem,liu2022erace,wu2019kepler,wu2018fuze}.
For instance, SemFuzz~\cite{you2017semfuzz} utilizes various vulnerability-related data including descriptions, source code, and patches to generate PoCs for kernel vulnerabilities autonomously.
However, the construction of a vulnerable kernel environment often receives insufficient attention, complicating the process of vulnerability reproduction.
To address this gap, this study is dedicated to \textit{identifying challenges and automating the generation of vulnerable kernel environments}.
We aim to simplify the process of reproducing kernel vulnerabilities, further reducing the time required to evaluate the risk and collect runtime attack behaviors associated with various kernel vulnerabilities. 

%% file: sections/03-motivations.tex
\subsection{Challenges}
\label{sec:motivations}

{\noindent \bf Motivating Example.} We discuss the challenges faced by kernel-vulnerable environment construction with an example of the real-world kernel vulnerability, \texttt{CVE-2021-22555}~\cite{nistCVE202122555}, an out-of-bounds issue in the Netfilter subsystem with the potential for local privilege escalation.

NVD's vulnerability record for CVE-2021-22555 delineates the affected kernel versions as those up to, but not including, v5.12~\cite{nistCVE202122555}. 
After cross-checking with the official Linux kernel release list~\cite{kernelIndexpublinuxkernel}, it appears that versions up to and including v5.11.22 are vulnerable to CVE-2021-22555, in line with NVD's assertion, implying that we can select any version from this range for vulnerability reproduction.

To reproduce this kernel vulnerability, we first compile the v5.11.22 kernel to create the vulnerable environment, then we test it with the public PoC~\cite{poccve202122555}, and it fails.
This failure is attributed to the presence of the patch for CVE-2021-22555 in the v5.11.22 source code.
Specifically, the three red lines from the vulnerable function \verb|xt_compat_target_from_user| in \autoref{fig:cve-2021-22555-motiv} (a), which are deleted by the patch~\cite{CVE202122555Patch} in \autoref{fig:cve-2021-22555-motiv} (b), have already disappeared in the v5.11.22 source code~\cite{cve202122555code}.
Moreover, our investigation confirms that kernel versions from v5.11.15 to v5.11.22 have similarly been patched against this vulnerability as well.
Consequently, the correct upper boundary for vulnerable kernels is v5.11.14 (inclusive).
Reproduction tests conducted on the v5.11.14 kernel using the public PoC confirm its susceptibility to CVE-2021-22555.
Therefore, attempting reproduction with v5.11.22 kernels would lead to an ineffective allocation of resources and time.

Following the determination of the vulnerable kernel version, the subsequent phase involves enabling the appropriate kernel configs to ensure the vulnerability is available and accessible in target kernel before initiating the kernel build.
In terms of CVE-2021-22555, our analysis of a segment of its patch~\cite{CVE202122555Patch}, as illustrated in \autoref{fig:cve-2021-22555-motiv} (b), identifies crucial configs in the Makefiles along the path leading to the inclusion of \verb|net/netfilter/x_tables.c|. These include \verb|CONFIG_NETFILTER| and \verb|CONFIG_NETFILTER_XTABLES| in \autoref{fig:cve-2021-22555-motiv} (c), as well as two other configs, \verb|CONFIG_NET| and \verb|CONFIG_INET|, upon which the former two configs depend, as specified in the Kconfig files in \autoref{fig:cve-2021-22555-motiv} (d). Furthermore, within the vulnerable file itself in \autoref{fig:cve-2021-22555-motiv} (a), a conditional compilation directive (\verb|#ifdef|) is observed, controlling the inclusion of the vulnerable function based on \verb|CONFIG_COMPAT|. By conducting a thorough manual analysis of all files affected by the patch in this manner, we can compile a comprehensive list of configs \textbf{necessary} to ensure the presence of the vulnerability in the target kernel version. The detailed analysis results for CVE-2021-22555, including the essential configs, are presented in \autoref{fig:cve-2021-22555-motiv} (e).

Unfortunately, even with the heuristic analysis above, the PoC for CVE-2021-22555~\cite{poccve202122555} remains non-functional. The issue stems from the PoC's specification of \verb|NFQUEUE| as the target in Netfilter, depicted in a code snippet from the PoC in \autoref{fig:cve-2021-22555-motiv} (f). This additionally requires the activation of \verb|CONFIG_NETFILTER_XT_TARGET_NFQUEUE| to support the specified target. Ultimately, enabling this config allows the PoC to trigger the vulnerability successfully. 

We identify two challenges that arise from the reproduction process of CVE-2021-22555:

\vspace{0.1cm}
\noindent \textbf{C1: Inaccurate Vulnerability Versions in CVE databases.}
The version range documented in vulnerability databases is not always accurate, requiring cross-checking with the kernel release list and the source code repository to identify a vulnerable version, which is quite time-consuming.

\vspace{0.1cm}
\noindent \textbf{C2: Non-obvious Vulnerability Configs.}
Vulnerability descriptions and reports often lack detailed information regarding the necessary configs that would ensure both the presence and user space accessibility of the vulnerability. Analysts have to download the source code of a vulnerable version, then compile the kernel using default configs (\verb|make defconfig|) and execute the PoC program to ascertain if it either crashes the system or triggers a Kernel Address SANitizer (KASAN) report. 

\begin{figure}[t]
\centering
\includegraphics[width=0.92\linewidth]{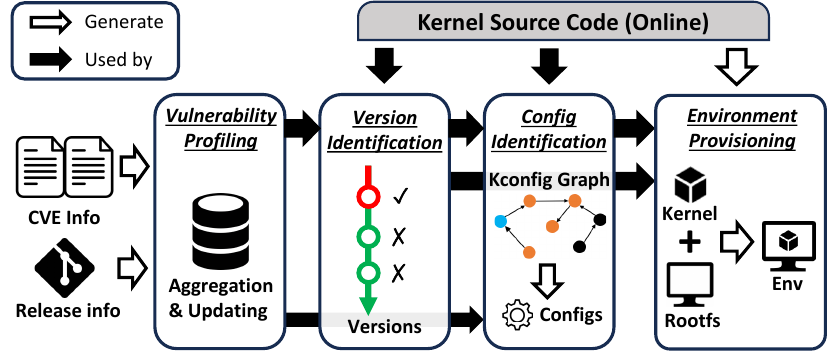}
\caption{Overview of \kjc Architecture}
\label{fig:appro-overview}
\end{figure}

This approach is generally effective for vulnerabilities located in the kernel's core functionalities, which are usually enabled by default configs. However, for numerous vulnerabilities tied to non-default functionalities or intricate subsystems, such as CVE-2021-22555,  relying solely on default configs proves inadequate for their activation. A manual analytical approach is typically needed to identify additional required configs. Alternatively, a heuristic analysis of the vulnerable code is often used alongside an examination of the Kbuild~\cite{kernelKconfigLanguage} and Kconfig~\cite{kernelKbuildx2014} mechanisms. Despite its monolithic nature, the kernel source code is hierarchically organized and developed modularly. The Kconfig and Kbuild systems work in tandem to tailor the kernel's functionalities and facilitate its build process. Statistically, the latest Linux kernel encompasses over 15,000 config items (\verb|CONFIG_*|), the majority of which are instrumental in deciding whether to incorporate specific source code files and activate particular features. Typically, kernel vulnerabilities manifest at the code block level. Thus, pinpointing the essential configs for introducing a vulnerability entails selecting the relevant configs from these 15,000 items, ensuring the inclusion of files and code blocks associated with the vulnerability. Given the extensive number of items and their intricate interrelations, manually undertaking this task is notably labor-intensive. What's worse, some configs needed by the reproduction are determined by analyzing the PoC programs, which are not always readily available, particularly for vulnerabilities that have only recently been disclosed.

%% file: sections/04-approach.tex
\section{Approach \label{sec:approach}}

\subsection{Overview}

\autoref{fig:appro-overview} delineates the architectural framework of \kjc, which is segmented into four distinct components: 
(1) a \textbf{vulnerability profiling} component, dedicated to the continuous aggregation and progressive updating of data concerning kernel vulnerabilities;
(2) a \textbf{vulnerability version identification} component, tasked with pinpointing an actual vulnerable kernel version corresponding to a specific CVE ID; 
(3) a \textbf{vulnerability config identification} component, focused on determining the requisite kernel configs for activating a particular vulnerability; and
(4) an \textbf{environment provisioning} component, instrumental in constructing the target kernel and establishing the comprehensive environment necessary for reproducing the vulnerability.

\vspace{0.1cm}
\noindent \textbf{Vulnerability Profiling.} \kjc is executed routinely to gather kernel vulnerability information for profiling each vulnerability.
The vulnerability profiles encompass essential information about vulnerabilities, patches, and kernel versions sourced from online CVE databases and official kernel repositories for subsequent analysis.
We will detail the vulnerability profiling component in \autoref{sec:appro-info-aggre}.

\vspace{0.1cm}
\noindent \textbf{Vulnerability Version Identification.} Upon receiving a CVE ID, \kjc validates the claimed version range's accuracy and identifies an authentically vulnerable version. 
To do so, \kjc ascertains the alleged vulnerable version ranges, correlated patches, and the implicated source code files to scrutinize the presence of patches within the purportedly vulnerable versions. \kjc designates a version as a false positive if a patch is detected within it. Subsequently, a retrospective examination is conducted through the kernel release chronology until a version devoid of the patch is located.
The vulnerability version identification component is delineated in \autoref{sec:appro-version-selection}.

\vspace{0.1cm}
\noindent \textbf{Vulneability Config Identification.} Once \kjc identifies the actual vulnerable kernel version, it proceeds to identify the specific kernel configs to activate the vulnerability. 
To streamline this intricate analysis, we propose a graph-based approach. 
The fundamental principle is to derive intuitive \textit{direct} configs from the vulnerability profiling component, build a Kconfig graph for the target kernel source code, and identify \textit{hidden} configs holding special relations with the \textit{direct} configs in the graph. 
The \textit{direct} and \textit{hidden} configs serve as the ultimate identification result. The vulnerability config identification component is detailed in \autoref{sec:appro-config-iden}.

\vspace{0.1cm}
\noindent \textbf{Environment Provisioning.} After \kjc discerns the kernel configs essential for the vulnerability, it integrates these configs with the foundational ones (\verb|defconfig|). Following this integration, \kjc proceeds to compile the kernel's source code of the identified vulnerable version, to build the kernel image. This constructed image, in conjunction with a root filesystem (rootfs), is then utilized to provision a virtual machine as the final reproduction environment atop hypervisors. 

\begin{figure}[t]
\centering
\includegraphics[width=0.92\linewidth]{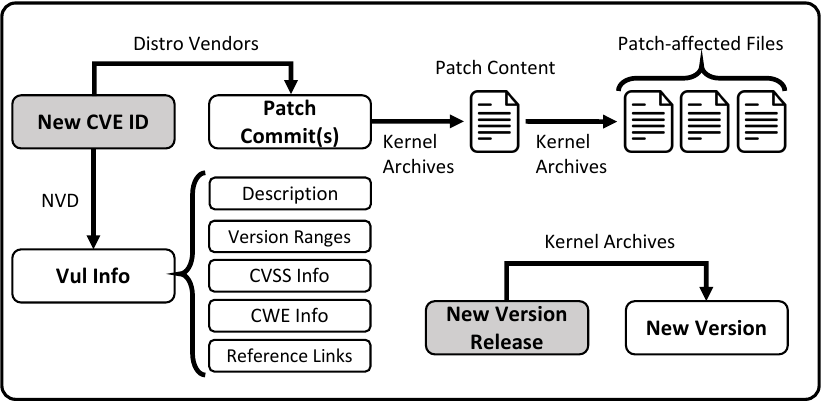}
\caption{Vulnerability Profiling}
\label{fig:appro-cve-profiling}
\end{figure}

\subsection{Vulnerability Profiling \label{sec:appro-info-aggre}}

\autoref{fig:appro-cve-profiling} illustrates our methodology for the incremental profiling of kernel vulnerabilities. Drawing from our hands-on experience in reproducing vulnerabilities, we identify three pivotal categories of important information:

\vspace{0.1cm}
\noindent \textbf{Vulnerability Information.} Given a CVE ID, the preliminary step involves comprehending its scope through its description. In rare cases, the description also includes the kernel configs needed to trigger the vulnerability~\cite{nistCVE201718344}. The stated range of vulnerable versions aids in targeting a specific kernel version, although it can occasionally be inaccurate, as exemplified in \autoref{sec:motivations}. Metrics such as CVSS and Common Weakness Enumeration (CWE) offer an initial severity assessment. Additionally, references and reports provide analyses from other experts.

\vspace{0.1cm}
\noindent \textbf{Patch Information.} Patches are invaluable in offering spatial and temporal insights into vulnerabilities. They assist in pinpointing the affected files and functions, understanding the root cause, and deriving strategies for triggering and exploitation. Patches are also instrumental in verifying the vulnerability of a targeted version and deriving the necessary kernel configs, providing pivotal information for reproduction.

\vspace{0.1cm}
\noindent \textbf{Release Version Information.} Maintaining an updated list of release versions streamlines the selection of an appropriate version for vulnerability reproduction. The version list is mainly used to fill the vulnerable version range used in \autoref{sec:appro-version-selection}.

When a new CVE of Linux kernel is reported, \kjc's initial step is to gather CVE details from NVD. Subsequently, it obtains patch commit IDs from distribution vendors (such as Ubuntu, Red Hat, and SUSE) for in-depth analysis. It is noteworthy that while NVD provides patch information for certain kernel vulnerabilities, the extent of this data is often surpassed by that from the vendors. Moreover, the occasional mislabeling of bug introduction commits as `Patch' in NVD~\cite{nistCVE202122555} complicates the identification of the actual patch link, leading us to prioritize vendor sources over NVD for patch information. Concurrently, our approach involves periodically reviewing new kernel releases, and updating the local list with any newly released versions.

\begin{figure}[t]
\centering
\includegraphics[width=0.92\linewidth]{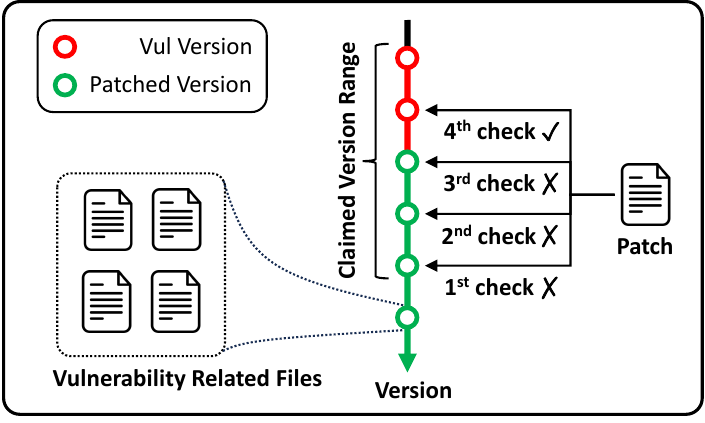}
\caption{Vulnerability Version Identification}
\label{fig:appro-version-selection}
\end{figure}

\subsection{Vulnerability Version Identification \label{sec:appro-version-selection}}

To detect spurious version range claims and select a real vulnerable kernel version, we propose a patch-based approach, shown in \autoref{fig:appro-version-selection}.
At a high level, \kjc inspects whether the corresponding patch to the vulnerability has been applied, by checking kernel versions in chronological order.

\kjc initially maps the claimed range of vulnerable versions with the list of kernel version releases. It then proceeds to verify the presence of the patch in each version, starting from the upper boundary of the inclusive version range and moving downwards. This process continues until it reaches the first version where the patch is not found.
Specifically, \kjc attempts to apply the patch to each kernel version.
If the version is already patched, a "re-patch" will be detected when applying the same patch again.
By doing so, \kjc identifies the first version where the patch is not found (i.e., the version without re-patching happens when applying the patch).
The version without a patch likely indicates the existence of the specific vulnerability.

\begin{figure}[t]
    \centering
    \includegraphics[width=0.92\linewidth]{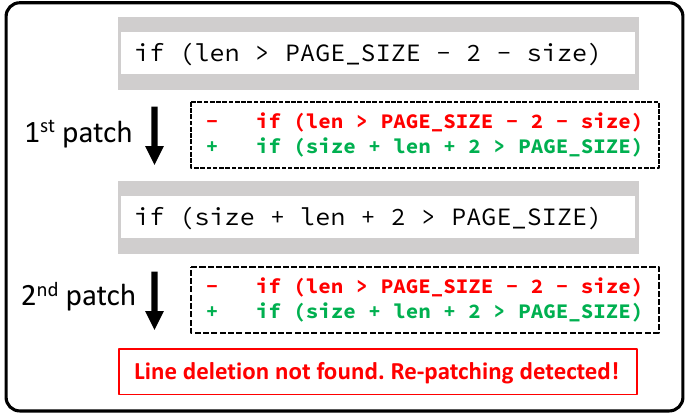}
    \caption{Detection of Re-patching for CVE-2022-0185. The grey rectangles around the target line represent the context.}
    \label{fig:repatch-demo}
\end{figure}

\autoref{fig:repatch-demo} employs the patch~\cite{CVE20220185Patch} for kernel vulnerability CVE-2022-0185~\cite{nistCVE20220185} as an exemplar to elucidate the process of re-patching detection. 
This specific patch incorporates both a line deletion and a line addition, framed within a 3-line context (represented by the grey rectangles around the target line in \autoref{fig:repatch-demo}). The first patching action substitutes the original \verb|if| statement with a new one, thereby remedying the vulnerability. Consequently, a subsequent patch attempt is unsuccessful, as the target line for deletion is no longer present following context alignment. This indicates that the source code has already undergone patching.

Essentially, the patch commit(s) content is derived from the \verb|diff| output for affected files in kernel source code of old \& new versions~\cite{kernelSubmittingPatches}. The finest granularity of patches is at the line level, with differences being delineated on a per-line basis, \textit{e.g.,} line additions or line deletions. The patch for a Linux kernel vulnerability could be composed of one or more commits; each commit may contain \verb|diff| results for one or more files; each file's \verb|diff| result contains either line additions, line deletions, or both. However, only the record of line modifications is inadequate for patching. Location information should also be provided to locate the accurate position to be patched. Due to the activity of Linux kernel development, locating via line numbers is prone to be erroneous when other commits occur in the same file. Linux kernel community adopts a \textit{N}-line context way~\cite{kernelApplyingPatches} to provide location information, i.e., attaching \textit{N} lines before and after the modified lines as context into the patch, which can tolerate the aforementioned line shift phenomenon to some extent. Although sometimes it would still fail (\textit{e.g.,} when both the \textit{N}-line context and the affected line(s) are replicated in the same file), this approach is easy to implement and efficient.

\begin{figure}[t]
\centering
\includegraphics[width=0.92\linewidth]{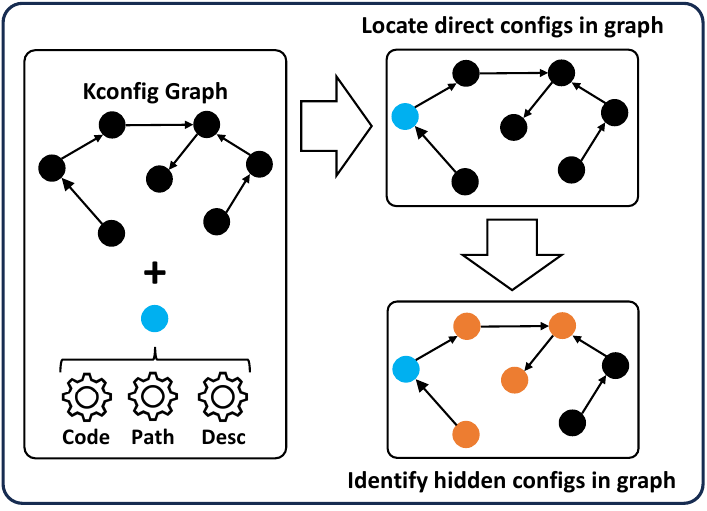}
\caption{Vulnerability Config Identification. Black dots represent all the configs in the Kconfig graph. Blue dots represent the set of \textit{direct configs}. Orange dots represent the set of \textit{hidden configs}. The three gears individually represent the set of code-level, path-level, and description-level configs.}
\label{fig:appro-config-identify}
\end{figure}

\subsection{Vulnerability Config Identification \label{sec:appro-config-iden}}

The vulnerability config identification process is composed of three stages: direct config identification, Kconfig graph construction, and hidden config identification, as shown by the pseudo-code in \autoref{algo:config-part} (the complete algorithm is in \autoref{sec:app-config-iden}) and \autoref{fig:appro-config-identify}.

Firstly, \kjc gathers the direct configs (\dc), composed of the description-level configs, the path-level configs (\dpc), and the code-level configs (\dcc), by analyzing the vulnerability description, patch, vulnerable source code, and Makefile(s). \ddc is derived from the description text, if any. \dpc represents the set of configs responsible for including the directory path towards the vulnerable file(s). \dcc represents the set of configs responsible for including the vulnerable code. \dc serves as the starting point for the graph analysis later on.

Secondly, \kjc starts from the root Kconfig file in the kernel source code, and recursively parses and imports new Kconfig files in sub-directories till leaf directories, to build the Kconfig graph (\graph). 
The specification of this directed graph is presented in \autoref{tab:config-kg}. There are three types of config definition in the Kconfig mechanism: \verb|config|, \verb|menuconfig|, and inner \verb|config| within blocks (\textit{e.g.,} \verb|choice| block), which are represented by vertexes in our graph.
Additionally, \verb|menu| is a special item in Kconfig, which also has relations with other configs, but it is not a real config.
\kjc takes it as a virtual vertex in the graph to ensure the connectivity and filters it out after getting all the hidden configs (\hc).
Besides vertexes, four types of directed edges are defined in the graph, which are derived from the specification of the Kconfig mechanism.
For example, as shown in \autoref{sec:motivations}, the \verb|menu "Core Netfilter Configurations"| \textit{depends on} the following three configs: \verb|CONFIG_NET|, \verb|CONFIG_INET|, and \verb|CONFIG_NETFILTER|.
As a result, there will be three \verb|depend| edges from the virtual vertex \verb|"Core Netfilter Configurations"| to the three config vertexes in the graph. Similarly, \verb|CONFIG_INET| has an \verb|opaque_depend| edge to \verb|CONFIG_NET|, as the former config is defined in the \verb|if| block of the latter one. The \verb|select| and \verb|imply| edges represent the relations in Kconfig that when config A is enabled, it will \verb|select| or \verb|imply| config B to be enabled as well, if there is a \verb|select CONFIG_B| statement within the definition of \verb|CONFIG_A|.

\input{tables/config-graph}

\begin{figure}[t]
\centering
\includegraphics[width=0.92\linewidth]{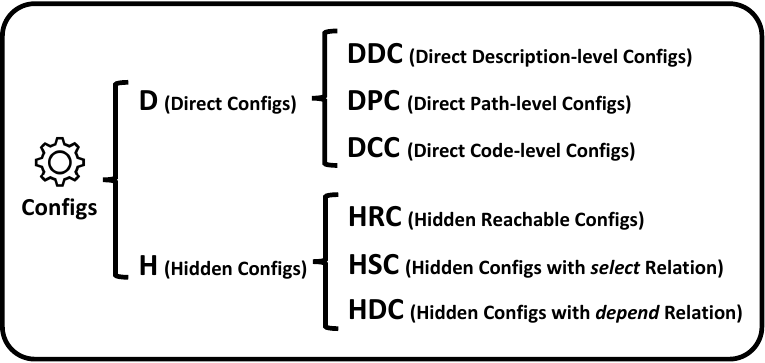}
\caption{Config Hierarchy and Categories}
\label{fig:eval-config-categories}
\end{figure}

Thirdly, \kjc leverages the direct configs and the Kconfig graph to get the hidden configs, which includes three types of configs (vertexes in the graph): (1) configs that are recursively reachable \textbf{from} any direct config (\hrc), (2) configs holding \textbf{one-hop} \verb|select| relation \textbf{to} any direct config (\hsc), and (3) configs holding \textbf{one-hop} \verb|depend| relation \textbf{to} any direct config (\hdc). As edges in the Kconfig graph are directed, the reachability of \hrc serves as a strong attribute to identify the vulnerability configs, which is consequently handled recursively. Nonetheless, \hsc and \hdc hold inverse relations to direct configs, which are only used to fulfill the config requirement for functionalities in complex subsystems, so \kjc conducts one-hop discovery to avoid introducing too many weakly related configs. Lastly, the set $S$, composed of both direct configs and hidden configs, serves as the final result for the vulnerability config identification process. It is noticeable that virtual vertexes are not considered as one hop, as they do not represent real configs, and will be filtered out from $S$. In summary, the identified configs can be categorized into six categories, as shown in \autoref{fig:eval-config-categories}.

\input{algorithms/vuln-config-identification}

\begin{figure}[t]
    \centering
    \includegraphics[width=\linewidth]{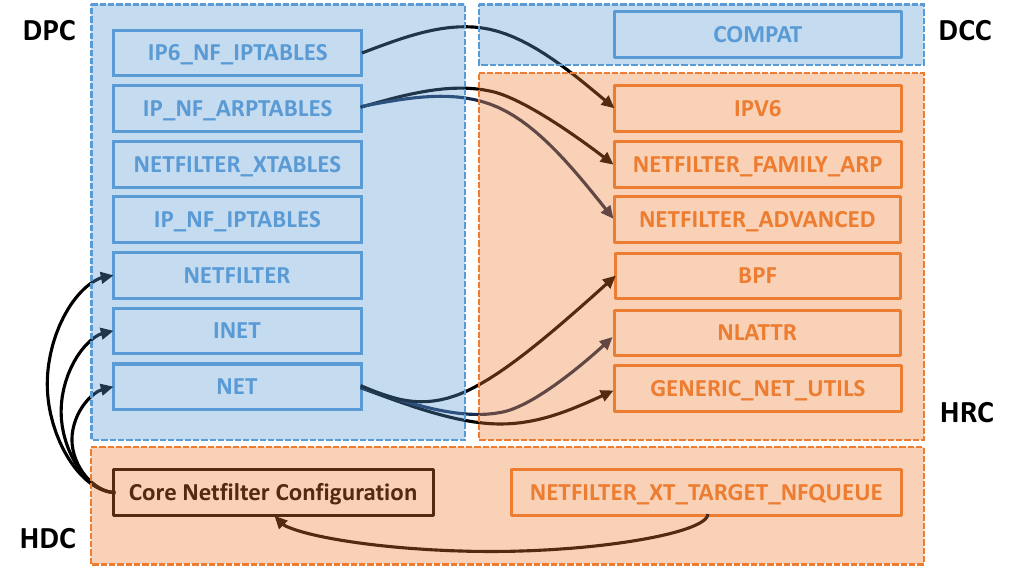}
    \caption{Graph-based Analysis for CVE-2021-22555. Blue boxes represent \textit{direct configs}, orange boxes represent \textit{hidden configs}, and the black box represents a menu item, which is to be filtered out at last. The \lstinline{CONFIG_} prefix is omitted.}
    \label{fig:cve-2021-22555-graph-res}
\end{figure}

For CVE-2021-22555, the heuristic analysis result in \autoref{fig:cve-2021-22555-motiv} (e) is equal to the combination of \dpc, \dcc and \hrc derived by the graph-based approach. Especially, the last pivotal config to activate this vulnerability, \verb|CONFIG_NETFILTER_XT_TARGET_NFQUEUE|, is successfully captured in \hdc. Therefore, this approach can provide adequate configs to support the reproduction of CVE-2021-22555, illustrated by the result graph shown in \autoref{fig:cve-2021-22555-graph-res}. Furthermore, \autoref{sec:eval-config-analysis} will analyze the contributions of configs in different sets (\ddc, \dpc, \dcc, \hrc, \hsc and \hdc) to the reproduction of representative kernel vulnerabilities. 

The approach to identifying necessary kernel configs for a specific vulnerability is derived from two observations: (1) to include a vulnerability, the vulnerable code must be activated by conditional compilation like \verb|#ifdef| (code-level configs), and the directory path towards the vulnerable file(s) must be included in Makefile(s) along the path (path-level configs); (2) the code-level and path-level configs have multiple necessary relations with other configs in the hierarchical Kconfig mechanism. Respectively, we regard configs in (1) as \textit{direct configs}, in that they are intuitive and can be found out by analyzing the patch and source code, and the configs in (2) as \textit{hidden configs}, because they hide in the complex structure of Kconfig and have to be identified with deep analysis. As a result, our initial idea to identify necessary configs is to first figure out all code-level (\dcc) and path-level (\dpc) configs, build a directed graph for all kernel configs and their relations from Kconfig files in the vulnerable source code, and use these configs as starting points to discover all reachable configs as hidden configs (\hrc).

However, only configs above do not necessarily ensure the activation of one vulnerability. There are exceptions in the reproduction experiences. For example, to trigger CVE-2017-18344~\cite{nistCVE201718344}, the PoC must access a pseudo-file, \verb|/proc/pid/timers|, which is only available when \verb|CONFIG_CHECKPOINT_RESTORE| is set~\cite{man7Proc5Linux}. For this case, the CVE description in NVD rarely provides the needed config items. Besides, for vulnerabilities located in complicated subsystems (\textit{e.g.,} Netfilter), special configs within the subsystems are needed to ensure the normal functionality in user space, illustrated by the CVE-2021-22555 example in \autoref{sec:motivations}. Consequently, \ddc, \hsc, and \hdc are proposed to improve the identification result. Considering the complexity of Linux Kconfig mechanism, we only derive \hsc and \hdc from  \textbf{one-hop} relations, instead of two or more hops, to avoid involving too many useless configs. This strategy proves to be effective and adequate in \autoref{sec:eval-config-analysis}.

%% file: tables/config-graph.tex

\begin{table}[t]
  \centering
  \setlength{\belowcaptionskip}{+10pt}
  \caption{Kconfig Graph Specification}
  \resizebox{\linewidth}{!}{%
  \begin{tabular}{lll}
    \hline
    \textbf{Element}  & \textbf{Type} & \textbf{Description} \\
    \hline
    config & vertex & "config" <symbol> in Kconfig \\
    menuconfig & vertex & "menuconfig" <symbol> in Kconfig \\
    inner\_config & vertex & configs embedded in other blocks in Kconfig \\
    menu & virtual vertex & "menu" in Kconfig \\
    depend & edge & "depends on <expr>" in Kconfig \\
    opaque\_depend & edge & statements like "if" and "source" in Kconfig \\
    select & edge & "select <symbol>" in Kconfig \\
    imply & edge & "imply <symbol>" in Kconfig \\
    \hline
  \end{tabular}
  }
  \label{tab:config-kg}
\end{table}

%% file: algorithms/vuln-config-identification.tex
\begin{algorithm}[t]
\caption{Vulnerability Config Identification}
\label{algo:config-part}
\SetKwComment{Comment}{/* }{ */}

\KwInput{$V$: Vulnerability Description; $P$: Patch Text for $V$; $SC$: Vulnerable Kernel Source Code}
\KwOutput{$S$: Set of Identified Config}

\vspace{0.5em}

\Proc{\textsc{GetVulConfigs}{$(V, P, SC)$}}{
    $D = \textsc{GetDirectConfigs}(V, P, SC)$\;
    $G = \textsc{BuildKconfigGraph}(SC)$\;
    $H = \textsc{GetHiddenConfigs}(D, G)$\;
    $S = D \cup H$\;
    \KwRet $S$\;
}
\vspace{0.5em}

\Proc{\textsc{GetDirectConfigs}{$(V, P, SC)$}}{
    $DF =$ affected files mentioned in $V$ (optional)\;
    $DDC =$ configs mentioned in $V$ (optional)\;
    $DPC =$ configs in $SC$ to enable $DF$ and files in $P$\;
    $DCC =$ configs in $SC$ to enable code in $P$ by \#ifdef\;
    \KwRet $DDC \cup DPC \cup DCC$\;
}
\vspace{0.5em}

\Proc{\textsc{BuildKconfigGraph}{$(SC)$}}{
    $G = $ empty graph\;
    $RK = $ root Kconfig file in $SC$\;
    \textit{AddKconfigNodes}($G, RK, SC$)\;
    \textit{AddKconfigEdges}($G, RK, SC$)\;
    \KwRet $G$\;
}
\vspace{0.5em}

\Proc{\textsc{GetHiddenConfigs}{$(D, G)$}}{
    $H = \emptyset$\;
    \ForEach{$c$ in $D$} {
        $HRC = $ reachable configs from $c$ in $G$\;
        $HSC = $ configs with $select$ relation to $c$ in $G$\;
        $HDC = $ configs with $depend$ relation to $c$ in $G$\; 
        Add $HRC, HSC, HDC$ into $H$\;
    }
    \KwRet $H$\;
}
\end{algorithm}

%% file: sections/05-implementation.tex
\section{Implementation \label{sec:impl}}

We develop \kjc in 3.4K lines of Python code. The key technical aspects of \kjc are detailed below.

\vspace{0.1cm}
\noindent \textbf{Patch Processing.} Patches play a crucial role in identifying both the presence of vulnerabilities and the necessary configs. \kjc integrates crawlers for both the Ubuntu Security Tracker~\cite{ubuntuSecurityUbuntu} and Red Hat Bugzilla~\cite{redhatBugzillaMain}, aimed at collecting patch commit IDs related to kernel vulnerabilities. When required, \kjc utilizes these IDs to fetch the raw patch content from the official kernel repository~\cite{kernelKernelorgRepositories}. The system then parses this data to pinpoint the modified files and specific line changes. These files are subsequently retrieved for applying patches and executing a direct config search. For identifying configs, \kjc inspects each vulnerable function's end where modifications occur, tracing back to the file's beginning to detect any conditional compilation directives (\verb|#ifdef CONFIG_*|).

\vspace{0.1cm}
\noindent \textbf{Kconfig Graph Construction.} The creation of a Kconfig graph is intricate due to the complex relations within the Linux kernel's Kconfig system. For example, a \verb|source "net/packet/Kconfig"| line within an \verb|if NET| condition implies that all configs in the \verb|net/packet/Kconfig| file are dependent on \verb|CONFIG_NET|. \kjc tackles this by applying a recursive strategy to form graph edges that represent these relations. Technologically, the tool utilizes NetworkX~\cite{networkxNetworkXx2014} to construct the graph and support traversal analysis.

\vspace{0.1cm}
\noindent \textbf{Rootfs and Virtual Machine Setup.} \kjc employs an adapted version of the \verb|create-image| script from the syzkaller project~\cite{githubSyzkallertoolscreateimageshMaster} for generating the base image for rootfs. In constructing each vulnerable environment, \kjc uses QEMU's~\cite{qemuoverlay} copy-on-write overlay feature to quickly establish rootfs for the targeted virtual machine (VM). Finally, the tool engages QEMU to launch the virtual machine, facilitating the vulnerability reproduction process. 

\vspace{0.1cm}
\noindent \textbf{User-friendly Interface.} The final phase in vulnerability reproduction is the testing of the developed PoC within the execution environment. This traditionally involves a cumbersome process where the analyst must transfer the PoC into the environment, execute it, and then monitor its behavior. Success on the initial attempt is rare, leading to a repetitive cycle of development and debugging of the PoC until it functions as intended. To streamline this often arduous task, \kjc's command-line interface has been thoughtfully designed to resemble Docker's user-friendly interface. It includes commands like \verb|cp| for copying files between the host and the VM, \verb|exec| for executing commands in the VM, and \verb|attach| for interacting with the VM, all aimed at simplifying the delivery and execution of the PoC within this VM.

%% file: sections/06-evaluation.tex
\section{Evaluation \label{sec:evaluation}}

\input{tables/eval-part-repro}

In this section, we evaluate \kjc by answering the following research questions (RQs):

\noindent \textbf{RQ1:} How is \kjc's performance in the reproduction of kernel vulnerabilities? (\autoref{sec:eval-effectiveness})

\noindent \textbf{RQ2:} How well do the configs identified by \kjc facilitate the reproduction of kernel vulnerabilities? (\autoref{sec:eval-config-analysis})

\noindent \textbf{RQ3:} How many false positive version claims in NVD can \kjc detect for Linux kernel vulnerabilities? (\autoref{sec:eval-fp-claim})

All the experiments are performed on a server with Intel(R) Xeon(R) CPU E5-2640 v4 @ 2.40GHz and 128 GB physical memory. The host OS is Ubuntu 22.04.3 LTS. Experiments for vulnerabilities from 2019 are conducted directly on the host. For vulnerabilities before 2019, we spawn a Ubuntu 18.04 container with Docker on this host for experiments, which avoids the incompatibility between the new compiler on the host and the old kernel source code.

\subsection{Dataset\label{sec:dataset}}

\begin{figure}[t]
    \centering
    \includegraphics[width=0.8\linewidth]{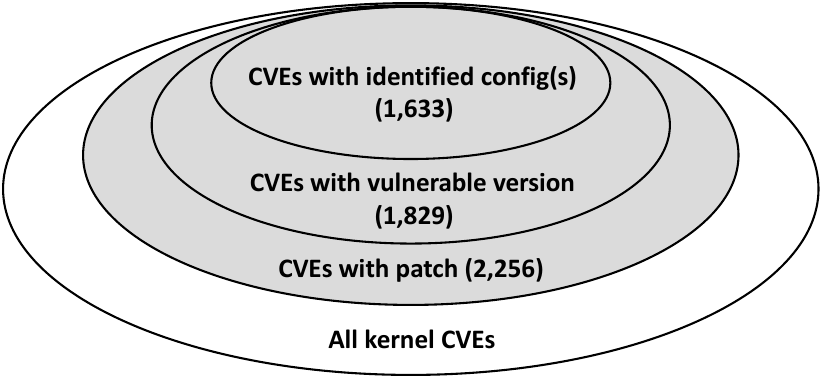}
    \caption{Construction of the Whole Dataset}
    \label{fig:whole-dataset}
\end{figure}

To adequately investigate the efficacy of our approaches, we create a comprehensive dataset of kernel vulnerabilities encompassing as many vulnerabilities as possible. We collect the CVE data (sourced from NVD) and the corresponding patch commits (from Linux distribution vendors, such as Ubuntu and Red Hat) for kernel vulnerabilities to date. After filtering out the invalid CVE IDs and those without patch commits recorded, we get 2,256 vulnerabilities.
The specifics of this data collection are elaborated in \autoref{sec:appro-info-aggre}. The 2,256 vulnerabilities serve as the whole dataset, among which we successfully identify the vulnerable versions for 1,829 vulnerabilities, and identify config(s) for 1,633 vulnerabilities, as shown in \autoref{fig:whole-dataset}. It should be noted that the version identification process aims to verify the correctness of existing version ranges, which is only done for kernel vulnerabilities whose vulnerable version ranges have already been claimed in NVD database. As a result, old CVEs without explicit version ranges in NVD are excluded. Furthermore, the absence of configs for a particular vulnerability does not imply \kjc’s inability to identify the necessary configs, as specific kernel vulnerabilities' availability and accessibility are not contingent on any kernel config.

\input{tables/eval-part-config-analysis}

Although more vulnerabilities could lead to a better evaluation of \kjc, conducting reproduction for the entire dataset is impractical. We decided to select vulnerabilities that are representative and have received widespread attention.
Consequently, we meticulously curated a subset of 66 kernel vulnerabilities in prominent research publications (SHARD~\cite{abubakar2021shard}, Midas~\cite{bhattacharyya2022midas}, KOOBE~\cite{chen2020koobe}, ELOISE~\cite{chen2020systematic}, SLAKE~\cite{chen2019slake}, Razzer~\cite{jeong2019razzer}, SegFuzz~\cite{jeong2023segfuzz}, AEM~\cite{jiang2023aem}, Pspray~\cite{lee2023pspray}, Hybrid~\cite{li2023hybrid}, DirtyCred~\cite{lin2022dirtycred}, LinKRID~\cite{liu2022linkrid}, KRover~\cite{pitigalaarachchi2023krover}, AlphaExp~\cite{wang2023alphaexp}, PET~\cite{wang2023pet}, Kepler~\cite{wu2019kepler}, PAL~\cite{yoo2022kernel}, DDRace~\cite{yuanddrace}, K(H)eaps~\cite{zeng2022playing}, RetSpill~\cite{zeng2023retspill}, StateFuzz~\cite{zhao2022statefuzz}) from ACM CCS, IEEE S\&P, NDSS and USENIX Security conferences in the past five years, with their associated PoCs from the Internet. This subset spans over nine years and covers a wide array of vulnerability types in various Linux kernel subsystems. To our knowledge, this dataset is currently the most extensive collection of kernel vulnerabilities with workable PoCs. Significantly, this subset falls within the 1,829 vulnerabilities for which vulnerable versions were identified. 

\vspace{0.1cm}
\noindent \textbf{Dataset Availability.} To facilitate ongoing research in the field of Linux kernel vulnerabilities, we will open-source the comprehensive dataset, encompassing vulnerabilities, their respective vulnerable versions, associated configs, and operational PoCs.

\subsection{Performance in Reproduction\label{sec:eval-effectiveness}}

For each vulnerability in the 66-CVE dataset, we run \kjc to build the vulnerable environments and then compile and execute the associated PoC in the target environments for reproduction. For each vulnerability with the source code downloaded, we run \kjc twice to build two environments with different config identification approaches: (1) for the first one, \kjc builds the environment with the identified configs using the aforementioned graph-based approach, and (2) for the second one, \kjc builds it with the default configs (\verb|make defconfig|) as baseline. The PoC will be executed in both of these two environments. After the PoC is executed, we observe whether a KASAN report is generated in the kernel log or the special OS state described in the PoC file is achieved, to determine the result of reproduction.

We conduct extra reproduction experiments for each vulnerability with false positive version claims detected to ensure the correctness of the version identification process. Specifically, we run \kjc to build a reproduction environment with a detected false positive version and \kjc-identified configs (a combination of identified non-default configs and the default configs from \verb|make defconfig|) for each case and execute the PoC within it.

Vulnerabilities that can only be reproduced with the activation of \kjc-identified configs or have false positive version claims detected by \kjc are listed in \autoref{tab:eval-part-repro}. The remaining results can be found in \autoref{tab:appendix-repro} of \autoref{sec:app-eval-res}. Results show that \kjc accomplishes the reproduction for all 66 vulnerabilities, indicating that it has effectively created a vulnerable environment for each of them. 

Among these vulnerabilities, 32 of 66 (48.5\%) need non-default configs identified by \kjc to be activated, and 34 of 66 (51.5\%) can be activated by default configs. Additionally, 4 of 66 (6.1\%) are detected to have false positive version claims in NVD. Interestingly, 2 of the 4 false positive cases (CVE-2021-22555 and CVE-2021-3573) are from the 32 vulnerabilities relying on non-default configs, and the other 2 vulnerabilities (CVE-2020-14381 and CVE-2020-25656) can be activated by default configs. As shown in \autoref{tab:eval-part-repro}, in summary, 34 of 66 (51.5\%) vulnerabilities can not be intuitively reproduced with claimed versions and default configs, due to either false positive version claims in NVD or extra non-default configs to activate the vulnerabilities. Among the 34 vulnerabilities, the ``kernel/bpf'' and ``net/netfilter'' subsystems contribute the largest number of vulnerabilities, which is 6 for both of them, compared with other subsystems in the table.

\subsection{Role of Identified Configs \label{sec:eval-config-analysis}}

As shown in \autoref{fig:eval-config-categories}, the vulnerability configs identified by \kjc are categorized into six categories: configs in vulnerability descriptions (\ddc), path-level configs (\dpc), code-level configs (\dcc), configs recursively reachable from direct configs (\hrc), configs with a one-hop \verb|select| relationship to any direct config (\hsc), and configs with a one-hop \verb|depend| relationship to any direct config (\hdc). \kjc can efficiently identify \ddc, \dpc, \dcc, \hrc, and \hsc through patch parsing and graph analysis. Comparatively, as demonstrated in \autoref{sec:motivations}, the discovery of \hsc and \hdc is more challenging but essential to activating vulnerabilities in complex subsystems. 
To this end, we analyze the 32 vulnerabilities relying on non-default configs, categorizing the identified configs and examining the relevance and contribution of \hsc and \hdc categories.

Vulnerabilities that rely on \hsc or \hdc are presented in \autoref{tab:eval-part-config-analysis}, and results for the remaining vulnerabilities can be found in \autoref{tab:appendix-config-analysis} of \autoref{sec:app-eval-res}. The results indicate that half of these vulnerabilities (16 out of 32) necessitate \hsc or \hdc for activation. Consequently, \hsc and \hdc identified by \kjc play an important role in constructing effective reproduction environments for kernel vulnerabilities.

It should be noted that not all of the identified \hsc and \hdc contribute to the activation of a specific vulnerability. For some vulnerabilities, especially those from the Netfilter subsystem, the number of exploited \hdc on average tends to be 280, due to the intricate dependencies of the subsystem, while only a small set of these configs actually matters. The effective \hdc for these vulnerabilities are analyzed in \autoref{sec:case-bpf-netfilter}. This phenomenon confirms the necessity of the one-hop design for \hsc and \hdc identification. However, we argue that such config identification results do not affect the efficacy of \kjc, as demonstrated by the reproduction results in \autoref{tab:eval-part-repro}. Besides, according to \autoref{tab:appendix-config-analysis}, on average, the Kconfig graph built by \kjc has 14,824 vertexes and 47,950 edges, which highlights the complexity of kernel and implies time-consuming work to handle the configs manually. Compared to the scale of this graph, the 280 configs introduced in \hdc are trivial.

Additionally, only in the case of CVE-2017-18344 were configs identified in the vulnerability description (\ddc), suggesting that descriptions are less impactful than other sources, such as patches and source code, in the config identification process. This fact also implies that it is infeasible for analysts to directly figure out the necessary configs for reproduction by retrieving the vulnerability descriptions.

\input{tables/eval-fp-cases}

\subsection{Detection of False Positive Version Claims \label{sec:eval-fp-claim}}

The accurate selection of vulnerable versions is crucial for the effective reproduction of kernel vulnerabilities. 
Wrong version can mislead analysts and result in significant time wastage. Thus, we aim to assess the proficiency of \kjc in identifying false positive version range claims in the NVD database for upstream Linux kernel vulnerabilities.

The whole CVE dataset mentioned in \autoref{sec:dataset} is used for this evaluation. \kjc's approach involves mapping the claimed vulnerable version ranges into a list of kernel release versions, followed by a downward analysis starting from the upper boundary to identify false positives. This process continues until a genuine vulnerable version is found or the lower boundary is reached, with each false positive instance recorded. Lastly, \kjc compiles and reports the total count of false positive versions detected.

The findings reveal that \kjc identifies false positive version range claims for 128 kernel vulnerabilities within the NVD database. The aggregate count of false positive versions is 3,042, averaging 24 false positive versions per identified vulnerability. The comprehensive table of these identification results (\autoref{tab:appendix-fp-cases}) is accessible in \autoref{sec:app-eval-res}, with the top 10 vulnerabilities sorted by false positive version count presented in \autoref{tab:eval-fp-cases}.

\subsection{Case Studies}

\subsubsection{Vulnerabilities in BPF \& Netfilter Subsystems \label{sec:case-bpf-netfilter}}

Within the evaluation results presented in \autoref{tab:appendix-repro}, we observe that two subsystems, eBPF and Netfilter, occur more often than other subsystems, each contributing six vulnerabilities to the dataset of 66 CVEs. 
Typically, more complex systems are susceptible to a higher number of vulnerabilities. 
In turn, the prevalence of vulnerabilities in eBPF and Netfilter underscores the complexity of these subsystems, a conclusion further corroborated in \autoref{sec:eval-config-analysis}. Analysis of config-dependent vulnerabilities (\autoref{tab:appendix-config-analysis}) reveals that most vulnerabilities associated with \hsc or \hdc configs originate from these two subsystems, indicating the increased challenges in manually constructing reproduction environments, while \kjc succeeds in identifying these configs and generating the environments for these vulnerabilities.

\input{tables/eval-case-study-bpf-netfilter}

For example, although the basic \verb|CONFIG_BPF| (\dpc) necessitated by the six eBPF vulnerabilities (CVE-2016-4557, CVE-2017-16995, CVE-2020-27194, CVE-2020-8835, CVE-2021-34866, and CVE-2021-3490), as indicated by file dependencies in the Makefile, is easy to figure out with manual analysis, the examination of the six reveals a consistent requirement for the \verb|CONFIG_BPF_SYSCALL| (\hsc identified by \kjc) to be enabled for their activation, which is comparatively non-intuitive. The absence of \verb|CONFIG_BPF_SYSCALL| will lead to wasted time and failed reproduction.

The analysis of the six Netfilter vulnerabilities reveals varied config requirements for each, along with their associated PoCs, complicating the manual construction of reproduction environments. Utilizing \kjc's identification results, we conducted a manual analysis to determine the minimal config set of \hsc and \hdc for each vulnerability. Our findings indicate that the effective hidden configs for these vulnerabilities are exclusively derived from their respective \hdc sets, as detailed in \autoref{tab:netfilter-config}. Specifically, we observed that the official announcement for CVE-2022-32250~\cite{seclistsCVE202232250} only cited the \verb|CONFIG_NETFILTER| and \verb|CONFIG_NF_TABLES| configs as necessary for activating this vulnerability. However, our reproduction demonstrated the additional requirement of \verb|CONFIG_NF_TABLES_IPV4|. Similarly, the disclosure for CVE-2023-32233~\cite{openwallOsssecurityCVE202332233} did not include the essential \verb|CONFIG_NFT_LOG| and \verb|CONFIG_NFT_QUOTA| config. The absence of these configs, as identified by \kjc, results in the failure of the original PoCs provided in these announcements during our reproduction experiments.

\subsubsection{Cross-minor-version FP: CVE-2018-1000028 \label{sec:case-2018-1000028}}

Within the comprehensive list of FP results detailed in \autoref{sec:app-eval-res}, CVE-2018-1000028~\cite{nistCVE20181000028} is particularly noteworthy. This vulnerability, stemming from incorrect access control in the kernel's NFS server implementation, potentially allows remote attackers to read or write files via NFS inappropriately. 

Our analysis highlights that CVE-2018-1000028 is the sole vulnerability within this list characterized by both a high CVSS severity score (7.4) and FP version ranges spanning two minor Linux release versions: NVD claims the vulnerable version range spans from v4.14.8 (inclusive) to v4.14.23 (inclusive) and from v4.15.1 (inclusive) to v4.15.7 (inclusive). However, our detection indicates that patches were applied from v4.14.16 (inclusive) to v4.14.23 (inclusive) and from v4.15.1 (inclusive) to v4.15.7 (inclusive), thereby reducing the scope of vulnerability. Drawing upon this analysis, we successfully reproduced the vulnerability in the v4.14.15 kernel version. Conversely, our attempts to reproduce it in the v4.14.16 kernel version were unsuccessful. 

For vulnerabilities like CVE-2018-1000028 that have incorrect version claims spanning more than one minor version number (\textit{e.g.,} v4.14 and v4.15 for CVE-2018-1000028) of kernel release versions in online databases, the impact is even more serious, as it is more difficult for analysts to select a vulnerable version for effective reproduction accurately. In such cases, \kjc can help save much time and effort on the pre-reproduction work.

%% file: tables/eval-part-repro.tex
\begin{table*}[ht]
    \centering
    \setlength{\belowcaptionskip}{+10pt}
    \caption{Vulnerability Reproduction Results. "RwKC?" and "RwDC?" denote reproducibility with \kjc-identified and default configs, respectively. "FPV?" indicates false positive version claims in NVD. Abbreviations: UAF = Use after Free; OOB = Out of Bounds; TOCTOU = Time of Check to Time of Use; DF = Double Free; ND = Null-pointer Dereference.}
    \resizebox{\linewidth}{!}{%
    \begin{tabular}{lllllllll}
      \hline
      \textbf{ID}  & \textbf{Type} & \textbf{CVSS} & \textbf{Subsystem} & \textbf{RwKC?} & \textbf{RwDC?} & \textbf{FPV?} & \textbf{Paper(s)} \\
      \hline
  CVE-2016-10150 & UAF   & 9.8 & virt/kvm   & \checkmark & \xmark  & \xmark & K(H)eaps, ELOISE, SLAKE, Kepler \\
  CVE-2016-4557 & UAF   & 7.8 & kernel/bpf   & \checkmark & \xmark  & \xmark & AEM, K(H)eaps, ELOISE, SLAKE, Kepler, RetSpill  \\
  CVE-2016-6187 & OOB   & 7.8 & security/apparmor    & \checkmark & \xmark  & \xmark & Pspray, PET, AEM, K(H)eaps, ELOISE, KOOBE, SLAKE, Kepler, RetSpill  \\
  CVE-2017-16995 & Logic     & 7.8  & kernel/bpf    & \checkmark & \xmark  & \xmark & AEM, Kepler \\
  CVE-2017-18344 & OOB   & 5.5 & kernel/time    & \checkmark & \xmark & \xmark & PET, AEM \\
  CVE-2017-2636 & DF    & 7.0  &  drivers/tty    & \checkmark & \xmark & \xmark & SegFuzz, PET, AEM, K(H)eaps, ExpRace, ELOISE, SLAKE, Kepler, Razzer, RetSpill  \\
  CVE-2017-6074 & DF    & 7.8 & net/dccp    & \checkmark & \xmark & \xmark & Pspray, AEM, K(H)eaps, ELOISE, SLAKE, Kepler, RetSpill  \\
  CVE-2017-8824 & UAF   & 7.8 & net/dccp    & \checkmark & \xmark & \xmark & PET, AEM, K(H)eaps, SLAKE, Kepler, RetSpill  \\
  CVE-2018-12233 & OOB   & 7.8 & fs/jfs   & \checkmark & \xmark & \xmark & ELOISE \\
  CVE-2018-5333 & ND    & 5.5  & net/rds & \checkmark & \xmark & \xmark & AEM \\
  CVE-2018-6555 & UAF   & 7.8 &  net/irda  & \checkmark & \xmark & \xmark & Pspray, AlphaEXP, AEM, K(H)eaps, ELOISE, SLAKE, RetSpill  \\
  CVE-2019-6974 & UAF   & 8.1 & virt/kvm   & \checkmark & \xmark & \xmark & SegFuzz, ExpRace \\
  CVE-2020-14381 & UAF   & 7.8 &  futex    & \checkmark     & \checkmark   & \checkmark   & AlphaEXP \\
  CVE-2020-16119 & UAF   & 7.8 & net/dccp & \checkmark & \xmark & \xmark & PET, DirtyCred \\
  CVE-2020-25656 & UAF   & 4.1 & drivers/tty      & \checkmark     & \checkmark   & \checkmark  & DDRace \\
  CVE-2020-25669 & UAF   & 7.8  & drivers/input  & \checkmark & \xmark & \xmark & StateFuzz \\
  CVE-2020-27194 & OOB   & 5.5  &  kernel/bpf   & \checkmark & \xmark & \xmark & AlphaEXP, DirtyCred \\
  CVE-2020-27830 & ND    & 5.5  & drivers/accessibility   & \checkmark & \xmark & \xmark & StateFuzz \\
  CVE-2020-28941 & ND    & 5.5 & drivers/accessibility    & \checkmark & \xmark & \xmark & StateFuzz \\
  CVE-2020-8835 & OOB   & 7.8 & kernel/bpf    & \checkmark & \xmark & \xmark & DirtyCred \\
  CVE-2021-22555 & OOB   & 7.8 & net/netfilter    & \checkmark & \xmark & \checkmark & PET, AlphaEXP, DirtyCred \\
  CVE-2021-26708 & UAF   & 7.0  &  net/vmw\_vsock    & \checkmark & \xmark & \xmark & AlphaEXP, DirtyCred \\
  CVE-2021-27365 & OOB   & 7.8 & drivers/scsi   & \checkmark & \xmark & \xmark & Hybrid, DirtyCred, RetSpill  \\
  CVE-2021-34866 & OOB   & 7.8  & kernel/bpf   & \checkmark & \xmark & \xmark & DirtyCred \\
  CVE-2021-3490 & OOB   & 7.8  & kernel/bpf   & \checkmark & \xmark & \xmark & DirtyCred, RetSpill  \\
  CVE-2021-3573 & UAF   & 6.4 & net/bluetooth   & \checkmark & \xmark & \checkmark & AlphaEXP \\
  CVE-2021-42008 & OOB   & 7.8  & drivers/net   & \checkmark & \xmark & \xmark & Hybrid, AlphaEXP, DirtyCred \\
  CVE-2021-43267 & OOB   & 9.8 & net/tipc   & \checkmark & \xmark & \xmark & Hybrid, AlphaEXP, KRover, DirtyCred, PET, RetSpill  \\
  CVE-2022-0995 & OOB   & 7.8 &  watch\_queue  & \checkmark & \xmark & \xmark & AlphaEXP, DirtyCred \\
  CVE-2022-1015 & OOB   & 6.6  & net/netfilter   & \checkmark & \xmark & \xmark & PET \\
  CVE-2022-25636 & OOB   & 7.8  & net/netfilter  & \checkmark & \xmark & \xmark & AlphaEXP, DirtyCred, RetSpill  \\
  CVE-2022-32250 & UAF   & 7.8 & net/netfilter   & \checkmark & \xmark & \xmark & Hybrid \\
  CVE-2022-34918 & OOB & 7.8 & net/netfilter  & \checkmark & \xmark  & \xmark  & PET, Hybrid \\
  CVE-2023-32233 & UAF & 7.8 &  net/netfilter  & \checkmark & \xmark & \xmark  & Hybrid \\
      \hline
    \end{tabular}
    }
    \label{tab:eval-part-repro}
  \end{table*}

%% file: tables/eval-part-config-analysis.tex
\begin{table*}[ht]
    \centering
    \setlength{\belowcaptionskip}{+10pt}
    \caption{Vulnerability Config Identification Statistics. The value in the Kernel column is the kernel source code version on which the configs are identified. Abbreviations are: configs in vulnerability descriptions (\ddc), path-level configs (\dpc), code-level configs (\dcc), configs that are reachable from any direct config (\hrc), configs holding one-hop select relation to any direct config (\hsc), configs holding one-hop depend relation to any direct config (\hdc). Numbers in red indicate that one or more configs in the column (\hsc/\hdc) are needed to activate the related vulnerability. }
    \resizebox{0.9\linewidth}{!}{%
    \begin{tabular}{llllllllll}
      \hline
    \textbf{CVE} & \textbf{Subsystem} & \textbf{Kernel} & \textbf{Kconfig Graph} & \textbf{DDC} & \textbf{DPC} & \textbf{DCC} & \textbf{HRC} & \textbf{HSC} & \textbf{HDC} \\
    \hline
    CVE-2016-10150 & virt/kvm & v4.8.12 & 12337v+38250e & 0     & 1     & 0     & 39    & 0     & \textcolor{red}{4} \\
    CVE-2016-4557 & kernel/bpf & v4.5.4 & 11845v+36556e & 0     & 1     & 0     & 0     & \textcolor{red}{2} & 0 \\
    CVE-2016-6187 & security/apparmor & v4.6.4 & 12021v+37152e & 0     & 1     & 0     & 14    & 0     & \textcolor{red}{2} \\
    CVE-2017-16995 & kernel/bpf & v4.14.8 & 13436v+41928e & 0     & 1     & 0     & 0     & \textcolor{red}{2} & 0 \\
    CVE-2019-6974 & virt/kvm & v4.20.7 & 14054v+44510e & 0     & 1     & 0     & 42    & 0     & \textcolor{red}{4} \\
    CVE-2020-27194 & kernel/bpf & v5.8.14 & 15340v+50234e & 0     & 1     & 0     & 0     & \textcolor{red}{2} & 1 \\
    CVE-2020-8835 & kernel/bpf & v5.6  & 14957v+48344e & 0     & 1     & 0     & 0     & \textcolor{red}{2} & 1 \\
    CVE-2021-22555 & net/netfilter & v5.11.14 & 15808v+51956e & 0     & 7     & 1     & 10    & 3 & \textcolor{red}{406} \\
    CVE-2021-34866 & kernel/bpf & v5.13.13 & 15982v+52565e & 0     & 1     & 0     & 0     & \textcolor{red}{2} & 3 \\
    CVE-2021-3490 & kernel/bpf & v5.12.3 & 15855v+52142e & 0     & 1     & 0     & 0     & \textcolor{red}{2} & 2 \\
    CVE-2021-3573 & net/bluetooth & v5.12.9 & 15851v+52148e & 0     & 1     & 0     & 32    & 0     & \textcolor{red}{45} \\
    CVE-2022-1015 & net/netfilter & v5.16.17 & 16244v+53665e & 0     & 1     & 0     & 4     & 0     & \textcolor{red}{241} \\
    CVE-2022-25636 & net/netfilter & v5.16.11 & 16243v+53663e & 0     & 4     & 0     & 19    & 2     & \textcolor{red}{241} \\
    CVE-2022-32250 & net/netfilter & v5.18.1 & 16542v+54754e & 0     & 1     & 0     & 4     & 0     & \textcolor{red}{238} \\
    CVE-2022-34918 & net/netfilter & v5.18.10 & 16548v+54770e & 0     & 1     & 0     & 4     & 0     & \textcolor{red}{238} \\
    CVE-2023-32233 & net/netfilter & v6.3.1 & 17120v+57166e & 0     & 2     & 0     & 5     & 0     & \textcolor{red}{317} \\
  \hline
    \end{tabular}%
    }
  \label{tab:eval-part-config-analysis}%
\end{table*}%

%% file: tables/eval-fp-cases.tex
\begin{table}[t]
  \centering
  \setlength{\belowcaptionskip}{+10pt}
  \caption{Vulnerabilities with FP Version Range Claims in NVD. Vulnerable Version is the first vulnerable version downawards adjacent to the lower boundary of the FP version range. FP Count is the number of FP versions within the FP version range. Abbreviation: FP = False Positive.}
  \resizebox{\linewidth}{!}{%
    \begin{tabular}{lllll}
    \hline
    \textbf{CVE}    & \textbf{CVSS} & \textbf{FP Version Range} & \textbf{Vulnerable Version} & \textbf{FP Count} \\
    \hline
    CVE-2017-1000407 & 7.4 & v4.14.6 -- v4.14.325 & v4.14.5 & 320 \\
    CVE-2017-18216 & 5.5 & v4.14.57 -- v4.14.325 & v4.14.56 & 269 \\
    CVE-2017-18224 & 4.7  & v4.14.57 -- v4.14.325 & v4.14.56 & 269 \\
    CVE-2020-35508 & 4.5  & v5.9.7 -- v5.11.22 & v5.9.6 & 229 \\
    CVE-2021-4002 & 4.4  & v5.15.5 -- v5.15.132 & v5.15.4 & 128 \\
    CVE-2021-4090 & 7.1  & v5.15.5 -- v5.15.132 & v5.15.4 & 128 \\
    CVE-2022-0264 & 5.5  & v5.15.11 -- v5.15.132 & v5.15.10 & 122 \\
    CVE-2021-4155 & 5.5  & v5.15.14 -- v5.15.132 & v5.15.13 & 119 \\
    CVE-2016-10906 & 7.0 & v4.4.191 -- v4.4.302 & v4.4.190 & 112 \\
    CVE-2015-4170 & 4.7 & v3.12.7 -- v3.13.3 & v3.12.6 & 72 \\
    \hline
    \end{tabular}%
    }
  \label{tab:eval-fp-cases}%
\end{table}%

%% file: tables/eval-case-study-bpf-netfilter.tex
\begin{table}[t]
  \centering
  \setlength{\belowcaptionskip}{+10pt}
  \caption{HDC Needed by Vulnerabilities in Netfilter Subsystem. The \lstinline{CONFIG_} prefix is omitted for clearness.}
  \resizebox{0.8\linewidth}{!}{%
    \begin{tabular}{ll}
\hline
\textbf{CVE}   & \textbf{HDC}                                                                        \\ \hline
CVE-2021-22555 & \lstinline{NETFILTER\_XT\_TARGET\_NFQUEUE}                                                                \\
CVE-2022-1015  & \lstinline{NF\_TABLES, NF\_TABLES\_IPV4}                                                                  \\
CVE-2022-25636 & \lstinline{NF\_TABLES, NF\_TABLES\_NETDEV}                                                                \\
CVE-2022-32250 & \lstinline{NF\_TABLES, NF\_TABLES\_IPV4}                                                                  \\
CVE-2022-34918 & \lstinline{NF\_TABLES, NF\_TABLES\_INET}                                                                  \\
\multirow{2}{*}{CVE-2023-32233} & \lstinline{NF\_TABLES, NF\_TABLES\_INET,}  \\
                                & \lstinline{NFT\_LOG, NFT\_QUOTA}           \\ \hline
\end{tabular}
  }
  \label{tab:netfilter-config}
\end{table}

%% file: sections/07-discussion.tex
\section{Discussion \label{sec:discussion}}

\subsection{Reproducibility and Exploitability}

Reproducibility and exploitability serve as vital metrics for gauging the severity of vulnerabilities and the associated risk in potentially susceptible environments. \kjc's methodology aids in the reproducibility assessment of Linux kernel vulnerabilities by creating an environment where the specific vulnerability is both present and accessible from user space. However, the environments generated by \kjc may not fulfill the prerequisites for successful exploitation. For instance, to reliably exploit vulnerabilities in a modern Linux kernel, where various mitigation techniques are in place, many exploitations utilize the Filesystem in Userspace (FUSE)~\cite{exploiterFUSELinux} or userfaultfd~\cite{synacktivExploitationDouble}. These mechanisms are contingent on the activation of non-default configs: \verb|CONFIG_FUSE_FS| for FUSE and \verb|CONFIG_USERFAULTFD| for userfaultfd. Since these configs are not related to the activation of vulnerabilities, their identification falls beyond the scope of this paper, which we intend to explore in future work.

\subsection{Dataset Preparation}

The 66-CVE dataset used in \autoref{sec:eval-effectiveness}, detailed in \autoref{tab:eval-part-repro} and more extensively in \autoref{tab:appendix-repro} of \autoref{sec:app-eval-res}, spans over nine years and encapsulates a wide array of vulnerability types within the Linux kernel's diverse subsystems. This comprehensive coverage underpins our assertion that the dataset is representative, and we anticipate that \kjc will demonstrate effective performance on other kernel vulnerabilities not explicitly covered in this paper. Time and labor constraints preclude a broader evaluation of \kjc against all existing vulnerabilities in upstream Linux kernels.
To facilitate future kernel vulnerability reproduction work, 
we plan to open-source \kjc and engage with the community for ongoing iterative evaluation and enhancement.

\subsection{Vulnerability Profiling}

Currently, \kjc leverages the NVD, which is in sync with the official MITRE CVE database~\cite{mitreCve}, to gather essential information necessary for reconstructing environments for Linux kernel vulnerabilities.
We acknowledge that if information is missing from these databases, the data collected may not suffice to accurately recreate such vulnerable environments.
However, the occurrence of absent information is exceedingly rare~\cite{dong2019towards} and does not significantly impact the generalizability or effectiveness of \kjc. Furthermore, as discussed in~\autoref{sec:appro-info-aggre}, \kjc is highly extensible.
It can swiftly respond to updates in CVE information, allowing it to quickly adapt and create the appropriate environment.

Additionally, the availability and quality of patches pose challenges to \kjc's ability to accurately identify the correct kernel version for reproducing vulnerable environments.
In terms of patch availability, our analysis of all CVEs affecting the upstream kernel~\cite{linuxkernelcvesLinuxKernel} reveals that patches are available for 95.7\% of vulnerabilities in upstream versions.
Regarding the quality of these patches, which is beyond the scope of our study of \kjc, one may consult existing literature~\cite{wu2023mitigating}, to understand how to extract information from existing patches.
To gather as much information on CVE and patches as possible for better profiling of kernel vulnerabilities, a viable approach involves sourcing data from the Linux kernel mailing list~\cite{lkmlLKMLORGLinux} or the kernel Bugzilla~\cite{kernelKernelorgBugzilla}.
This approach represents an engineering enhancement, and we aim to refine our vulnerability profiling methodology in future development and iterations of the open-source project of \kjc.

%% file: sections/08-related.tex
\section{Related Work \label{sec:related}}

\vspace{0.1cm}
\noindent \textbf{Reproducibility Assessment.} Mu \textit{et al.}~\cite{mu2018understanding} conducted the first empirical analysis encompassing both the vulnerable environment and the PoC program. Chen \textit{et al.}~\cite{chen2021robin} introduced a binary similarity-based method to deduce the specific build configurations for vulnerabilities in userland programs. Additionally, various studies~\cite{brumley2008automatic,cadar2006exe,you2017semfuzz} have investigated the automated generation of PoC using vulnerability-related data, such as descriptions, source code, and patches. Pham \textit{et al.}~\cite{pham2015hercules} utilized symbolic execution for generating PoC for binaries, with a notable focus by You \textit{et al.}~\cite{you2017semfuzz} on the automated generation of PoC for Linux kernel vulnerabilities. Avgerinos \textit{et al.}~\cite{Avgerinos2011AEGAE} pioneered the concept of Automatic Exploit Generation (AEG). Subsequent works~\cite{alhuzali2016chainsaw,bao2017your,heelan2019gollum,hu2015automatic,park2022fugio,wang2018revery,wang2021maze,zhang2023automated} have expanded AEG to various vulnerability types, including data-oriented, web-based, and heap-based vulnerabilities. Specifically, in the context of Linux kernel vulnerabilities, significant contributions~\cite{chen2020koobe,jiang2023aem,liu2022erace,wu2019kepler,wu2018fuze} have been made toward AEG for vulnerabilities of certain types.

\vspace{0.1cm}
\noindent \textbf{Vulnerability Version Assessment.} Constructing the vulnerable environment necessitates identifying a kernel version with known vulnerabilities. However, the accuracy of version information from online vulnerability databases is not always guaranteed. A viable method to ascertain the vulnerability of a specific kernel version involves examining whether the code manifesting the vulnerability or its associated patch exists within the target source code or the final compiled programs. Nguyen \textit{et al.}~\cite{nguyen2016automatic} employed a technique to detect vulnerable versions by verifying the presence of a known vulnerability's code in its preceding versions. Bao \textit{et al.}~\cite{bao2022v} introduced the V-SZZ algorithm, designed to validate vulnerable versions across 172 CVEs from 55 C/C++ or Java projects, by pinpointing the earliest commit that altered the lines of code in question. Further, Jiang \textit{et al.}~\cite{jiang2020pdiff} and Zhang \textit{et al.}~\cite{zhang2018precise} concentrated on testing the presence of patches in downstream kernel image binaries. Dong \textit{et al.}~\cite{dong2019towards} proposed VIEM, a model utilizing deep learning to identify discrepancies, such as those in vulnerable version listings, between the NVD database, unstructured CVE descriptions, and external reports. Nonetheless, these studies above do not specifically target the upstream versions of Linux kernel source code. While VIEM effectively detects inconsistencies, it does not discern which piece of conflicting information is accurate.

\vspace{0.1cm}
\noindent \textbf{Linux Kconfig Analysis.} Kernel configs represent another crucial element of the vulnerable environment. The complexity inherent in the Kconfig mechanism of modern kernels, characterized by an extensive array of configs, renders the identification of specific configs responsible for introducing vulnerabilities a non-trivial task. To date, no research directly addresses this specific challenge. However, several studies within the software engineering domain offer valuable insights into understanding and managing Kconfig complexities. Notable works include the analysis and remediation of Kconfig defects by Franz \textit{et al.}~\cite{franz2021configfix}, the evaluation of Kconfig models by Hengelein \textit{et al.}~\cite{hengelein2015analyzing}, the detection of configuration-related errors by Oh \textit{et al.}~\cite{oh2021finding}, and the lightweight extraction of variability information by Ruprecht \textit{et al.}~\cite{ruprecht2015lightweight}.

%% file: sections/09-conclusion.tex
\section{Conclusion \label{sec:conclusion}}

This paper introduced \kjc, an innovative tool designed to automate the generation of vulnerable environments for Linux kernel vulnerabilities. \kjc stands out for its patch-based vulnerable version detection capability, enabling it to discern and rectify erroneous version information in NVD. This functionality is instrumental in accurately identifying the genuinely vulnerable version of a given Linux kernel vulnerability. Additionally, \kjc leverages both direct and hidden config information within the Kconfig graph, ensuring the activation of the target vulnerability within the constructed environment. We evaluate \kjc with real-world kernel vulnerabilities collected from existing research. The evaluation results underscore the \kjc’s proficiency in accurately pinpointing the genuine vulnerable versions and requisite kernel configs. More importantly, \kjc demonstrates its capability to reliably establish vulnerable environments conducive to the reproduction of kernel vulnerabilities.

%% file: main.bbl

%% file: sections/99-appendix.tex
\clearpage
\appendix

\section{Automated Vulnerability Config Identifying \label{sec:app-config-iden}}

The entire algorithm of the graph-based approach for identifying necessary kernel configs is shown in \autoref{algo:config}. \textsc{GetVulConfigs} is the entrance of this algorithm.

\input{algorithms/vuln-config-iden-full}

\section{Additional Evaluation Results \label{sec:app-eval-res}}

\noindent \textbf{Performance in Vulnerability Reproduction.} The evaluation results for reproducing the 66 vulnerabilities are presented in \autoref{tab:appendix-repro}. Results indicate that \kjc effectively creates a vulnerable environment for all 66 vulnerabilities. Of the 66 vulnerabilities, 32 vulnerabilities (48.5\%) require non-default configs identified by \kjc to be activated, whereas 34 vulnerabilities (51.5\%) can be activated by default configs. Furthermore, 4 vulnerabilities (6.1\%) have been identified to have false positive version claims in NVD.

\input{tables/appendix-repro}

\vspace{0.1cm}
\noindent \textbf{Role of Configs Identified by \kjc.} The evaluation results for analyzing the role of identified vulnerability configs are presented in \autoref{tab:appendix-config-analysis}. Out of the 32 vulnerabilities, 16 require \hsc or \hdc configs to be activated. The Kconfig graph constructed by \kjc has an average of 14,824 vertices and 47,950 edges, which indicates the intricate nature of the Linux kernel. This complexity also implies that manual handling of the configs can be a time-consuming process.

\input{tables/appendix-config-analysis}


\vspace{0.1cm}
\noindent \textbf{Detection of False Positive Version Claims.} The evaluation results for false positive version detection on the 2,256 vulnerabilities are presented in \autoref{tab:appendix-fp-cases}. The findings reveal that \kjc identifies false positive version range claims in 128 out of 2,256 (5.7\%) kernel vulnerabilities within the NVD database. The aggregate count of false positive versions is 3,042, which averages to 24 false positive versions per identified vulnerability.

\input{tables/appendix-fp-cases}

%% file: algorithms/vuln-config-iden-full.tex
\removelatexerror
\begin{algorithm}[H]
\resizebox{0.85\linewidth}{!}{%
\begin{minipage}{\linewidth}
\caption{Vulnerability Config Identification}
\label{algo:config}
\SetKwComment{Comment}{/* }{ */}
\KwInput{$V$: Vulnerability Description; $P$: Patch Text for $V$; $SC$: Vulnerable Kernel Source Code}
\KwOutput{$S$: Set of Identified Configs}

\vspace{0.5em}

\Proc{\textsc{GetVulConfigs}{$(V, P, SC)$}}{
    $D = \textsc{GetDirectConfigs}(V, P, SC)$\;
    $G = \textsc{BuildKconfigGraph}(SC)$\;
    $H = \textsc{GetHiddenConfigs}(D, G)$\;
    $S = D \cup H$\;
    \KwRet $S$\;
}
\vspace{0.5em}

\Proc{\textsc{GetDirectConfigs}{$(V, P, SC)$}}{
    $DF =$ affected files mentioned in $V$ (optional)\;
    $DDC =$ configs mentioned in $V$ (optional)\;
    $DPC =$ configs in $SC$ to enable $DF$ and files in $P$\;
    $DCC =$ configs in $SC$ to enable code in $P$ by \#ifdef\;
    \KwRet $DDC \cup DPC \cup DCC$\;
}
\vspace{0.5em}

\Proc{\textsc{BuildKconfigGraph}{$(SC)$}}{
    $G = $ empty graph\;
    $RK = $ root Kconfig file in $SC$\;
    \textsc{AddKconfigNodes}($G, RK, SC$)\;
    \textsc{AddKconfigEdges}($G, RK, SC$)\;
    \KwRet $G$\;
}
\vspace{0.5em}

\Proc{\textsc{GetHiddenConfigs}{$(D, G)$}}{
    $H = \emptyset$\;
    \ForEach{$c$ in $D$}{
        $HRC = $ reachable configs from $c$ in $G$\;
        $HSC = $ configs with $select$ relation to $c$ in $G$\;
        $HDC = $ configs with $depend$ relation to $c$ in $G$\;
        Add $HRC, HSC, HDC$ into $H$\;
    }
    \KwRet $H$\;
}
\vspace{0.5em}

\Proc{\textsc{AddKconfigNodes}{$(G, K, SC)$}}{
    \ForEach{$line$ in $K$}{
        \If{$line$ defines config $c$}{
            Add $c$ into $G$\;
        }
        \If{$line$ imports new Kconfig file $nk$}{
            \textsc{AddKconfigNodes}($G, nk, SC$)\;
        }
    }
}
\vspace{0.5em}

\Proc{\textsc{AddKconfigEdges}{$(G, K, SC)$}}{
    \ForEach{$line$ in $K$}{
        \If{$line$ defines config $c$} {
            Add edges from/to $c$ into $G$\;
        }
        \If{$line$ defines any block from $(menu, if, choice)$} {
            Add edges derived from the block into $G$\;
        }
        \If{$line$ imports new Kconfig file $nk$} {
            \textsc{AddKconfigEdges}($G, nk, SC$)\;
        }
    }
}
\end{minipage}
}
\end{algorithm}

%% file: tables/appendix-repro.tex
\begin{table*}[ht]
    \centering
    \setlength{\belowcaptionskip}{+10pt}
    \caption{Vulnerability Reproduction Results. "RwKC?" and "RwDC?" denote reproducibility with \kjc-identified and default configs, respectively. "FPV?" indicates false positive version claims in NVD. Green CVE IDs signify the need for non-default configs or presence of false positive claims. Abbreviations: UAF = Use after Free; OOB = Out of Bounds; TOCTOU = Time of Check to Time of Use; DF = Double Free; ND = Null-pointer Dereference.}
    \resizebox{0.87\linewidth}{!}{%
    \begin{tabular}{lllllllll}
      \hline
      \textbf{CVE}  & \textbf{Type} & \textbf{CVSS} & \textbf{Subsystem} & \textbf{RwKC?} & \textbf{RwDC?} & \textbf{FPV?} & \textbf{Paper(s)} \\
      \hline
  CVE-2015-3636 & UAF   & 4.9   & net/ipv4     & \checkmark     & \checkmark   & \xmark   & SLAKE \\
  CVE-2016-0728 & UAF   & 7.8   & security/keys  & \checkmark     & \checkmark   & \xmark   & K(H)eaps, LinKRID, SHARD, ELOISE, SLAKE, RetSpill \\
  \textcolor{green}{CVE-2016-10150} & UAF   & 9.8 & virt/kvm   & \checkmark & \xmark  & \xmark & K(H)eaps, ELOISE, SLAKE, Kepler \\
  \textcolor{green}{CVE-2016-4557} & UAF   & 7.8 & kernel/bpf   & \checkmark & \xmark  & \xmark & AEM, K(H)eaps, ELOISE, SLAKE, Kepler, RetSpill  \\
  \textcolor{green}{CVE-2016-6187} & OOB   & 7.8 & security/apparmor    & \checkmark & \xmark  & \xmark & Pspray, PET, AEM, K(H)eaps, ELOISE, KOOBE, SLAKE, Kepler, RetSpill  \\
  CVE-2016-6516 & TOCTOU   & 7.4   & fs/ioctl   & \checkmark     & \checkmark    & \xmark  & Midas \\
  CVE-2016-8655 & UAF   & 7.8   & net/packet    & \checkmark     & \checkmark   & \xmark   & SegFuzz, AEM, K(H)eaps, ExpRace, SLAKE, Kepler, Razzer \\
  CVE-2016-9793 & OOB   & 7.8   & net/core     & \checkmark     & \checkmark    & \xmark  & AEM, Kepler \\
  CVE-2017-1000112 & OOB   & 7.0  & net/ipv4, net/ipv6        & \checkmark     & \checkmark    & \xmark  & AEM, ELOISE, KOOBE, SLAKE \\
  CVE-2017-10661 & UAF   & 7.0  &  fs/timefd       & \checkmark     & \checkmark    & \xmark  & AEM, K(H)eaps, ELOISE, SLAKE, Kepler, RetSpill  \\
  CVE-2017-11176 & UAF   & 7.8  &  ipc/mqueue    & \checkmark     & \checkmark    & \xmark  & AEM, K(H)eaps, RetSpill  \\
  CVE-2017-15265 & UAF   & 7.0  & sound/core       & \checkmark     & \checkmark   & \xmark   & ExpRace \\
  CVE-2017-15649 & UAF   & 7.8  & net/packet    & \checkmark     & \checkmark   & \xmark   & SegFuzz, AEM, K(H)eaps, ELOISE, SLAKE, Kepler \\
  \textcolor{green}{CVE-2017-16995} & Logic     & 7.8  & kernel/bpf    & \checkmark & \xmark  & \xmark & AEM, Kepler \\
  CVE-2017-17052 & UAF   & 7.8  & kernel/fork    & \checkmark     & \checkmark  & \xmark    & ELOISE, SLAKE \\
  CVE-2017-17053 & UAF   & 7.0  &  asm/mmu\_context       & \checkmark     & \checkmark   & \xmark   & ELOISE, SLAKE, Kepler \\
  CVE-2017-17712 & UAF   & 7.0  &  net/ipv4       & \checkmark     & \checkmark  & \xmark    & SegFuzz, ExpRace, Razzer \\
  \textcolor{green}{CVE-2017-18344} & OOB   & 5.5 & kernel/time    & \checkmark & \xmark & \xmark & PET, AEM \\
  \textcolor{green}{CVE-2017-2636} & DF    & 7.0  &  drivers/tty    & \checkmark & \xmark & \xmark & SegFuzz, PET, AEM, K(H)eaps, ExpRace, ELOISE, SLAKE, Kepler, Razzer, RetSpill  \\
  CVE-2017-5123 & Logic     & 8.8 & kernel/exit      & \checkmark     & \checkmark & \xmark     & Hybrid, AEM, SHARD, Kepler \\
  \textcolor{green}{CVE-2017-6074} & DF    & 7.8 & net/dccp    & \checkmark & \xmark & \xmark & Pspray, AEM, K(H)eaps, ELOISE, SLAKE, Kepler, RetSpill  \\
  CVE-2017-7184 & OOB   & 7.8 & net/xfrm       & \checkmark     & \checkmark  & \xmark    & Pspray, PET, AEM, K(H)eaps, ELOISE, KOOBE, SLAKE, Kepler, RetSpill  \\
  CVE-2017-7308 & OOB   & 7.8 & net/packet       & \checkmark     & \checkmark  & \xmark    & PET, AEM, K(H)eaps, PAL, SHARD, ELOISE, KOOBE, SLAKE, Kepler, RetSpill  \\
  CVE-2017-7533 & OOB   & 7.0   & fs/notify       & \checkmark     & \checkmark   & \xmark   & SegFuzz, Pspray, K(H)eaps, ExpRace, ELOISE, KOOBE, RetSpill  \\
  \textcolor{green}{CVE-2017-8824} & UAF   & 7.8 & net/dccp    & \checkmark & \xmark & \xmark & PET, AEM, K(H)eaps, SLAKE, Kepler, RetSpill  \\
  CVE-2017-8890 & DF    & 7.8  &  net/ipv4      & \checkmark     & \checkmark   & \xmark   & AEM, K(H)eaps, ELOISE, SLAKE, Kepler \\
  CVE-2018-10840 & OOB   & 6.6 &  fs/ext4   & \checkmark     & \checkmark   & \xmark   & SLAKE \\
  CVE-2018-12232 & ND    & 5.9  &  net/socket      & \checkmark     & \checkmark    & \xmark  & SegFuzz \\
  \textcolor{green}{CVE-2018-12233} & OOB   & 7.8 & fs/jfs   & \checkmark & \xmark & \xmark & ELOISE \\
  CVE-2018-18559 & UAF   & 8.1 &  net/packet  & \checkmark     & \checkmark & \xmark     & ELOISE, SLAKE \\
  \textcolor{green}{CVE-2018-5333} & ND    & 5.5  & net/rds & \checkmark & \xmark & \xmark & AEM \\
  \textcolor{green}{CVE-2018-6555} & UAF   & 7.8 &  net/irda  & \checkmark & \xmark & \xmark & Pspray, AlphaEXP, AEM, K(H)eaps, ELOISE, SLAKE, RetSpill  \\
  CVE-2019-15666 & UAF   & 4.4  & net/xfrm   & \checkmark     & \checkmark & \xmark     & AlphaEXP, AEM, DirtyCred \\
  \textcolor{green}{CVE-2019-6974} & UAF   & 8.1 & virt/kvm   & \checkmark & \xmark & \xmark & SegFuzz, ExpRace \\
  \textcolor{green}{CVE-2020-14381} & UAF   & 7.8 &  futex    & \checkmark     & \checkmark   & \checkmark   & AlphaEXP \\
  CVE-2020-14386 & OOB   & 7.8 & net/packet      & \checkmark     & \checkmark  & \xmark    & PET, DirtyCred \\
  \textcolor{green}{CVE-2020-16119} & UAF   & 7.8 & net/dccp & \checkmark & \xmark & \xmark & PET, DirtyCred \\
  \textcolor{green}{CVE-2020-25656} & UAF   & 4.1 & drivers/tty      & \checkmark     & \checkmark   & \checkmark  & DDRace \\
  \textcolor{green}{CVE-2020-25669} & UAF   & 7.8  & drivers/input  & \checkmark & \xmark & \xmark & StateFuzz \\
  \textcolor{green}{CVE-2020-27194} & OOB   & 5.5  &  kernel/bpf   & \checkmark & \xmark & \xmark & AlphaEXP, DirtyCred \\
  \textcolor{green}{CVE-2020-27830} & ND    & 5.5  & drivers/accessibility   & \checkmark & \xmark & \xmark & StateFuzz \\
  CVE-2020-28097 & OOB   & 5.9 & vgacon   & \checkmark     & \checkmark  & \xmark    & StateFuzz \\
  \textcolor{green}{CVE-2020-28941} & ND    & 5.5 & drivers/accessibility    & \checkmark & \xmark & \xmark & StateFuzz \\
  \textcolor{green}{CVE-2020-8835} & OOB   & 7.8 & kernel/bpf    & \checkmark & \xmark & \xmark & DirtyCred \\
  CVE-2021-20226 & UAF   & 7.8 & io\_uring    & \checkmark     & \checkmark  & \xmark    & AlphaEXP \\
  \textcolor{green}{CVE-2021-22555} & OOB   & 7.8 & net/netfilter    & \checkmark & \xmark & \checkmark & PET, AlphaEXP, DirtyCred \\
  CVE-2021-22600 & DF    & 7.0   & net/packet      & \checkmark     & \checkmark  & \xmark    & DirtyCred \\
  \textcolor{green}{CVE-2021-26708} & UAF   & 7.0  &  net/vmw\_vsock    & \checkmark & \xmark & \xmark & AlphaEXP, DirtyCred \\
  \textcolor{green}{CVE-2021-27365} & OOB   & 7.8 & drivers/scsi   & \checkmark & \xmark & \xmark & Hybrid, DirtyCred, RetSpill  \\
  CVE-2021-33909 & OOB   & 7.8 & fs/seq\_file.c     & \checkmark     & \checkmark  & \xmark    & AlphaEXP, DirtyCred \\
  \textcolor{green}{CVE-2021-34866} & OOB   & 7.8  & kernel/bpf   & \checkmark & \xmark & \xmark & DirtyCred \\
  \textcolor{green}{CVE-2021-3490} & OOB   & 7.8  & kernel/bpf   & \checkmark & \xmark & \xmark & DirtyCred, RetSpill  \\
  \textcolor{green}{CVE-2021-3573} & UAF   & 6.4 & net/bluetooth   & \checkmark & \xmark & \checkmark & AlphaEXP \\
  CVE-2021-41073 & UAF   & 7.8  & io\_uring     & \checkmark     & \checkmark  & \xmark    & AlphaEXP, DirtyCred \\
  CVE-2021-4154 & UAF   & 8.8 & kernel/cgroup      & \checkmark     & \checkmark & \xmark     & PET, DirtyCred, RetSpill  \\
  \textcolor{green}{CVE-2021-42008} & OOB   & 7.8  & drivers/net   & \checkmark & \xmark & \xmark & Hybrid, AlphaEXP, DirtyCred \\
  \textcolor{green}{CVE-2021-43267} & OOB   & 9.8 & net/tipc   & \checkmark & \xmark & \xmark & Hybrid, AlphaEXP, KRover, DirtyCred, PET, RetSpill  \\
  CVE-2022-0185 & OOB   & 8.4 & fs/fs\_context    & \checkmark     & \checkmark & \xmark     & PET, Hybrid, AlphaEXP, DirtyCred, RetSpill  \\
  \textcolor{green}{CVE-2022-0995} & OOB   & 7.8 &  watch\_queue  & \checkmark & \xmark & \xmark & AlphaEXP, DirtyCred \\
  \textcolor{green}{CVE-2022-1015} & OOB   & 6.6  & net/netfilter   & \checkmark & \xmark & \xmark & PET \\
  CVE-2022-1786 & UAF   & 7.8  & io\_uring  & \checkmark     & \checkmark   & \xmark   & Hybrid, RetSpill  \\
  CVE-2022-24122 & UAF   & 7.8  & kernel/ucount     & \checkmark     & \checkmark  & \xmark    & DirtyCred \\
  \textcolor{green}{CVE-2022-25636} & OOB   & 7.8  & net/netfilter  & \checkmark & \xmark & \xmark & AlphaEXP, DirtyCred, RetSpill  \\
  \textcolor{green}{CVE-2022-32250} & UAF   & 7.8 & net/netfilter   & \checkmark & \xmark & \xmark & Hybrid \\
  \textcolor{green}{CVE-2022-34918} & OOB & 7.8 & net/netfilter  & \checkmark & \xmark  & \xmark  & PET, Hybrid \\
  \textcolor{green}{CVE-2023-32233} & UAF & 7.8 &  net/netfilter  & \checkmark & \xmark & \xmark  & Hybrid \\
      \hline
    \end{tabular}
    }
    \label{tab:appendix-repro}
  \end{table*}

%% file: tables/appendix-config-analysis.tex
\begin{table*}[ht]
    \centering
    \setlength{\belowcaptionskip}{+10pt}
    \caption{Vulnerability Config Identification Statistics. The value in the Kernel column is the kernel source code version on which the configs are identified. Abbreviations are: configs in vulnerability descriptions (\ddc), path-level configs (\dpc), code-level configs (\dcc), configs that are reachable from any direct config (\hrc), configs holding one-hop select relation to any direct config (\hsc), configs holding one-hop depend relation to any direct config (\hdc). Numbers in red indicate that one or more configs in the column (\hsc/\hdc) are needed to activate the related vulnerability.}
    \resizebox{0.92\linewidth}{!}{%
    \begin{tabular}{llllllllll}
      \hline
    \textbf{CVE} & \textbf{Subsystem} & \textbf{Kernel} & \textbf{Kconfig Graph} & \textbf{DDC} & \textbf{DPC} & \textbf{DCC} & \textbf{HRC} & \textbf{HSC} & \textbf{HDC} \\
    \hline
    CVE-2016-10150 & virt/kvm & v4.8.12 & 12337v+38250e & 0     & 1     & 0     & 39    & 0     & \textcolor{red}{4} \\
    CVE-2016-4557 & kernel/bpf & v4.5.4 & 11845v+36556e & 0     & 1     & 0     & 0     & \textcolor{red}{2} & 0 \\
    CVE-2016-6187 & security/apparmor & v4.6.4 & 12021v+37152e & 0     & 1     & 0     & 14    & 0     & \textcolor{red}{2} \\
    CVE-2017-16995 & kernel/bpf & v4.14.8 & 13436v+41928e & 0     & 1     & 0     & 0     & \textcolor{red}{2} & 0 \\
    CVE-2017-18344 & kernel/time & v4.14.7 & 13436v+41929e & 2     & 0     & 0     & 3     & 0     & 3 \\
    CVE-2017-2636 & drivers/tty & v4.10.2 & 12706v+39472e & 0     & 1     & 0     & 17    & 0     & 0 \\
    CVE-2017-6074 & net/dccp & v4.9.12 & 12519v+38798e & 0     & 1     & 0     & 9     & 0     & 0 \\
    CVE-2017-8824 & net/dccp & v4.14.19 & 13441v+41941e & 0     & 1     & 0     & 9     & 0     & 0 \\
    CVE-2018-12233 & fs/jfs & v4.17.1 & 13588v+43147e & 0     & 1     & 0     & 4     & 0     & 4 \\
    CVE-2018-5333 & net/rds & v4.14.13 & 13437v+41930e & 0     & 1     & 0     & 9     & 0     & 3 \\
    CVE-2018-6555 & net/irda & v4.16.18 & 13626v+42836e & 0     & 2     & 1     & 7     & 0     & 37 \\
    CVE-2019-6974 & virt/kvm & v4.20.7 & 14054v+44510e & 0     & 1     & 0     & 42    & 0     & \textcolor{red}{4} \\
    CVE-2020-16119 & net/dccp & v5.8.10 & 15340v+50231e & 0     & 1     & 0     & 5     & 0     & 0 \\
    CVE-2020-25669 & drivers/input & v5.9.9 & 15456v+50684e & 0     & 3     & 0     & 3     & 37    & 3 \\
    CVE-2020-27194 & kernel/bpf & v5.8.14 & 15340v+50234e & 0     & 1     & 0     & 0     & \textcolor{red}{2} & 1 \\
    CVE-2020-27830 & drivers/accessibility & v5.9.13 & 15457v+50695e & 0     & 2     & 0     & 19    & 0     & 0 \\
    CVE-2020-28941 & drivers/accessibility & v5.9.9 & 15456v+50684e & 0     & 2     & 0     & 19    & 0     & 0 \\
    CVE-2020-8835 & kernel/bpf & v5.6  & 14957v+48344e & 0     & 1     & 0     & 0     & \textcolor{red}{2} & 1 \\
    CVE-2021-22555 & net/netfilter & v5.11.14 & 15808v+51956e & 0     & 7     & 1     & 10    & 3 & \textcolor{red}{406} \\
    CVE-2021-26708 & net/vmw\_vsock & v5.10.12 & 15674v+51410e & 0     & 1     & 0     & 4     & 0     & 6 \\
    CVE-2021-27365 & drivers/scsi & v5.11.3 & 15808v+51950e & 0     & 2     & 0     & 22    & 8     & 0 \\
    CVE-2021-34866 & kernel/bpf & v5.13.13 & 15982v+52565e & 0     & 1     & 0     & 0     & \textcolor{red}{2} & 3 \\
    CVE-2021-3490 & kernel/bpf & v5.12.3 & 15855v+52142e & 0     & 1     & 0     & 0     & \textcolor{red}{2} & 2 \\
    CVE-2021-3573 & net/bluetooth & v5.12.9 & 15851v+52148e & 0     & 1     & 0     & 32    & 0     & \textcolor{red}{45} \\
    CVE-2021-42008 & drivers/net & v5.13.12 & 15981v+52560e & 0     & 2     & 0     & 18    & 0     & 14 \\
    CVE-2021-43267 & net/tipc & v5.14.15 & 16009v+52674e & 0     & 1     & 0     & 5     & 0     & 4 \\
    CVE-2022-0995 & watch\_queue & v5.16.4 & 16244v+53661e & 0     & 1     & 1     & 0     & 0     & 1 \\
    CVE-2022-1015 & net/netfilter & v5.16.17 & 16244v+53665e & 0     & 1     & 0     & 4     & 0     & \textcolor{red}{241} \\
    CVE-2022-25636 & net/netfilter & v5.16.11 & 16243v+53663e & 0     & 4     & 0     & 19    & 2     & \textcolor{red}{241} \\
    CVE-2022-32250 & net/netfilter & v5.18.1 & 16542v+54754e & 0     & 1     & 0     & 4     & 0     & \textcolor{red}{238} \\
    CVE-2022-34918 & net/netfilter & v5.18.10 & 16548v+54770e & 0     & 1     & 0     & 4     & 0     & \textcolor{red}{238} \\
    CVE-2023-32233 & net/netfilter & v6.3.1 & 17120v+57166e & 0     & 2     & 0     & 5     & 0     & \textcolor{red}{317} \\
  \hline
    \end{tabular}%
    }
  \label{tab:appendix-config-analysis}%
\end{table*}%

%% file: tables/appendix-fp-cases.tex
\begin{table*}[ht]
    \centering
    \setlength{\belowcaptionskip}{+10pt}
    \caption{Vulnerabilities with FP Version Range Claims in NVD. Vulnerable Version is the first vulnerable version downawards adjacent to the lower boundary of the FP version range. FP Count is the number of FP versions within the FP version range. Abbreviation: FP = False Positive.}
    \resizebox{\linewidth}{!}{%
    \begin{tabular}{lllll|lllll}
    \hline
    \textbf{CVE} & \textbf{CVSS} & \textbf{FP Version Range} & \textbf{Vulnerable Version} & \textbf{FP Count} & \textbf{CVE} & \textbf{CVSS} & \textbf{FP Version Range} & \textbf{Vulnerable Version} & \textbf{FP Count} \\
    \hline 
    CVE-2012-4444 & 5.0     & v2.6.36 -- v2.6.36 & v2.6.35.14 & 1     & CVE-2021-3744 & 5.5   & v5.14.10 -- v5.14.21 & v5.14.9 & 12 \\
    CVE-2012-5375 & 4.0     & v3.8 -- v3.8 & v3.7.10 & 1     & CVE-2021-3753 & 4.7   & v5.14.1 -- v5.14.21 & v5.14 & 21 \\
    CVE-2012-6536 & 2.1   & v3.5.7 -- v3.5.7 & v3.5.6 & 1     & CVE-2021-3764 & 5.5   & v5.14.10 -- v5.14.20 & v5.14.9 & 11 \\
    CVE-2012-6538 & 1.9   & v3.5.7 -- v3.5.7 & v3.5.6 & 1     & CVE-2021-4002 & 4.4   & v5.15.5 -- v5.15.132 & v5.15.4 & 128 \\
    CVE-2012-6542 & 1.9   & v3.5.5 -- v3.5.7 & v3.5.4 & 3     & CVE-2021-4090 & 7.1   & v5.15.5 -- v5.15.132 & v5.15.4 & 128 \\
    CVE-2012-6545 & 1.9   & v3.5.5 -- v3.5.7 & v3.5.4 & 3     & CVE-2021-4155 & 5.5   & v5.15.14 -- v5.15.132 & v5.15.13 & 119 \\
    CVE-2012-6647 & 4.9   & v3.4.8 -- v3.4.9 & v3.4.7 & 2     & CVE-2021-4203 & 6.8   & v5.14.10 -- v5.14.21 & v5.14.9 & 12 \\
    CVE-2013-0217 & 5.2   & v3.7.8 -- v3.7.8 & v3.7.7 & 1     & CVE-2022-0264 & 5.5   & v5.15.11 -- v5.15.132 & v5.15.10 & 122 \\
    CVE-2013-3224 & 4.9   & v3.8.11 -- v3.9 & v3.8.10 & 4     & CVE-2022-0322 & 5.5   & v5.14.14 -- v5.14.21 & v5.14.13 & 8 \\
    CVE-2013-3236 & 4.9   & v3.9 -- v3.9 & v3.8.13 & 1     & CVE-2022-0494 & 4.4   & v5.16.13 -- v5.16.20 & v5.16.12 & 8 \\
    CVE-2014-0205 & 6.9   & v2.6.36.1 -- v2.6.36.4 & v2.6.36 & 4     & CVE-2022-1011 & 7.8   & v5.16.15 -- v5.16.20 & v5.16.14 & 6 \\
    CVE-2015-1593 & 5.0     & v3.18.9 -- v3.18.9 & v3.18.8 & 1     & CVE-2022-1055 & 7.8   & v5.16.6 -- v5.16.20 & v5.16.5 & 15 \\
    CVE-2015-4170 & 4.7   & v3.12.7 -- v3.13.3 & v3.12.6 & 72    & CVE-2022-1198 & 5.5   & v5.16.15 -- v5.16.20 & v5.16.14 & 6 \\
    CVE-2015-8944 & 5.5   & v4.6 -- v4.7 & v4.5.7 & 9     & CVE-2022-1263 & 5.5   & v5.17.3 -- v5.17.15 & v5.17.2 & 13 \\
    CVE-2016-10906 & 7.0     & v4.4.191 -- v4.4.302 & v4.4.190 & 112   & CVE-2022-1353 & 7.1   & v5.16.19 -- v5.16.20 & v5.16.18 & 2 \\
    CVE-2016-2085 & 5.5   & v4.4.2 -- v4.4.8 & v4.4.1 & 7     & CVE-2022-1419 & 7.8   & v5.5.5 -- v5.5.19 & v5.5.4 & 15 \\
    CVE-2016-2384 & 4.6   & v4.4.2 -- v4.4.8 & v4.4.1 & 7     & CVE-2022-1734 & 7.0     & v5.17.7 -- v5.17.15 & v5.17.6 & 9 \\
    CVE-2016-2550 & 5.5   & v4.4.4 -- v4.4.8 & v4.4.3 & 5     & CVE-2022-1852 & 5.5   & v5.18.2 -- v5.18.19 & v5.18.1 & 18 \\
    CVE-2016-5400 & 4.3   & v4.6.6 -- v4.6.6 & v4.6.5 & 1     & CVE-2022-2078 & 5.5   & v5.18.2 -- v5.18.19 & v5.18.1 & 18 \\
    CVE-2016-6156 & 5.1   & v4.6.6 -- v4.6.6 & v4.6.5 & 1     & CVE-2022-2318 & 5.5   & v5.18.10 -- v5.18.19 & v5.18.9 & 10 \\
    CVE-2016-9604 & 4.4   & v4.10.13 -- v4.11 & v4.10.12 & 6     & CVE-2022-2380 & 5.5   & v5.17.2 -- v5.17.15 & v5.17.1 & 14 \\
    CVE-2017-1000407 & 7.4   & v4.14.6 -- v4.14.325 & v4.14.5 & 320   & CVE-2022-2905 & 5.5   & v5.19.6 -- v5.19.17 & v5.19.5 & 12 \\
    CVE-2017-12762 & 9.8   & v4.12.5 -- v4.12.5 & v4.12.4 & 1     & CVE-2022-3078 & 5.5   & v5.17.2 -- v5.17.15 & v5.17.1 & 14 \\
    CVE-2017-16643 & 6.6   & v4.13.11 -- v4.13.11 & v4.13.10 & 1     & CVE-2022-3303 & 4.7   & v5.19.9 -- v5.19.17 & v5.19.8 & 9 \\
    CVE-2017-18216 & 5.5   & v4.14.57 -- v4.14.325 & v4.14.56 & 269   & CVE-2022-3543 & 5.5   & v6.0.3 -- v6.0.19 & v6.0.2 & 17 \\
    CVE-2017-18224 & 4.7   & v4.14.57 -- v4.14.325 & v4.14.56 & 269   & CVE-2022-3594 & 5.3   & v6.0.3 -- v6.0.19 & v6.0.2 & 17 \\
    CVE-2017-5669 & 7.8   & v4.10.2 -- v4.10.17 & v4.10.1 & 16    & CVE-2022-3595 & 5.5   & v6.0.16 -- v6.0.19 & v6.0.15 & 4 \\
    CVE-2017-5986 & 5.5   & v4.9.11 -- v4.9.11 & v4.9.10 & 1     & CVE-2022-3707 & 5.5   & v6.0.19 -- v6.0.19 & v6.0.18 & 1 \\
    CVE-2017-7518 & 7.8   & v4.11.8 -- v4.11.12 & v4.11.7 & 5     & CVE-2022-45869 & 5.5   & v6.0.11 -- v6.0.19 & v6.0.10 & 9 \\
    CVE-2017-7558 & 7.5   & v4.12.14 -- v4.13 & v4.12.13 & 2     & CVE-2022-48502 & 7.1   & v6.1.40 -- v6.1.54 & v6.1.39 & 15 \\
    CVE-2018-1000028 & 7.4   & v4.14.16 -- v4.14.23, v4.15.1 -- v4.15.7 & v4.14.15 & 15   & CVE-2023-0469 & 5.5   & v6.0.11 -- v6.0.19 & v6.0.10 & 9 \\
    CVE-2018-10853 & 7.8   & v4.17.2 -- v4.17.19 & v4.17.1 & 18    & CVE-2023-0590 & 4.7   & v6.0.6 -- v6.0.19 & v6.0.5 & 14 \\
    CVE-2018-1120 & 5.3   & v4.16.10 -- v4.16.18 & v4.16.9 & 9     & CVE-2023-0615 & 5.5   & v6.0.7 -- v6.1.54 & v6.0.6 & 68 \\
    CVE-2018-18281 & 7.8   & v4.18.16 -- v4.18.20 & v4.18.15 & 5     & CVE-2023-1079 & 6.8   & v6.2.3 -- v6.2.16 & v6.2.3 & 14 \\
    CVE-2019-12819 & 5.5   & v4.20.17 -- v4.20.17 & v4.20.16 & 1     & CVE-2023-1206 & 5.7   & v6.4.8 -- v6.4.16 & v6.4.7 & 9 \\
    CVE-2019-14814 & 7.8   & v5.2.17 -- v5.2.21 & v5.2.16 & 5     & CVE-2023-1249 & 5.5   & v5.17.2 -- v5.17.15 & v5.17.1 & 14 \\
    CVE-2019-18282 & 5.3   & v5.3.10 -- v5.3.10 & v5.3.9 & 1     & CVE-2023-1513 & 3.3   & v6.1.13 -- v6.1.54 & v6.1.12 & 42 \\
    CVE-2019-19036 & 5.5   & v5.3.4 -- v5.3.12 & v5.3.3 & 9     & CVE-2023-1990 & 4.7   & v6.2.8 -- v6.2.16 & v6.2.7 & 9 \\
    CVE-2019-3460 & 6.5   & v5.0.6 -- v5.1 & v5.0.5 & 17    & CVE-2023-1998 & 5.6   & v6.2.3 -- v6.2.16 & v6.2.2 & 14 \\
    CVE-2019-3901 & 4.7   & v4.5.6 -- v4.7.10 & v4.5.5 & 21    & CVE-2023-2002 & 6.8   & v6.3.1 -- v6.3.13 & v6.3  & 13 \\
    CVE-2020-10720 & 5.5   & v5.1.7 -- v5.1.21 & v5.1.6 & 15    & CVE-2023-2019 & 4.4   & v5.19.2 -- v5.19.17 & v5.19.1 & 16 \\
    CVE-2020-14314 & 5.5   & v5.8.4 -- v5.8.9 & v5.8.3 & 6     & CVE-2023-2162 & 5.5   & v6.1.11 -- v6.1.54 & v6.1.10 & 44 \\
    CVE-2020-14381 & 7.8   & v5.5.12 -- v5.5.19 & v5.5.11 & 8     & CVE-2023-2177 & 5.5   & v5.18.16 -- v5.18.19 & v5.18.15 & 4 \\
    CVE-2020-15436 & 6.7   & v5.7.6 -- v5.7.19 & v5.7.5 & 14    & CVE-2023-2598 & 7.8   & v6.3.2 -- v6.3.6 & v6.3.1 & 5 \\
    CVE-2020-15437 & 4.4   & v5.7.11 -- v5.7.19 & v5.7.10 & 9     & CVE-2023-2985 & 5.5   & v6.2.3 -- v6.2.16 & v6.2.2 & 14 \\
    CVE-2020-25641 & 5.5   & v5.8.8 -- v5.8.13 & v5.8.7 & 6     & CVE-2023-3111 & 7.8   & v5.19.4 -- v5.19.17 & v5.19.3 & 14 \\
    CVE-2020-25656 & 4.1   & v5.9.5 -- v5.9.16 & v5.9.4 & 12    & CVE-2023-3141 & 7.1   & v6.3.4 -- v6.3.6 & v6.3.3 & 3 \\
    CVE-2020-35508 & 4.5   & v5.9.7 -- v5.11.22 & v5.9.6 & 229   & CVE-2023-3159 & 6.7   & v5.17.7 -- v5.17.15 & v5.17.6 & 9 \\
    CVE-2021-20261 & 6.4   & v4.4.262 -- v4.4.302 & v4.4.261 & 41    & CVE-2023-3161 & 5.5   & v6.1.11 -- v6.1.54 & v6.1.10 & 44 \\
    CVE-2021-20320 & 5.5   & v5.14.7 -- v5.14.21 & v5.14.6 & 15    & CVE-2023-32254 & 8.1   & v6.3.2 -- v6.3.13 & v6.3.1 & 12 \\
    CVE-2021-20321 & 4.7   & v5.14.12 -- v5.14.21 & v5.14.11 & 10    & CVE-2023-32258 & 8.1   & v6.3.2 -- v6.3.9 & v6.3.1 & 8 \\
    CVE-2021-20322 & 7.4   & v5.14.4 -- v5.14.21 & v5.14.3 & 18    & CVE-2023-3268 & 7.1   & v6.3.2 -- v6.3.13 & v6.3.1 & 12 \\
    CVE-2021-22555 & 7.8   & v5.11.15 -- v5.11.22 & v5.11.14 & 8     & CVE-2023-3269 & 7.8   & v6.4.1 -- v6.4.16 & v6.4  & 16 \\
    CVE-2021-31916 & 6.7   & v5.11.11 -- v5.11.22 & v5.11.10 & 12    & CVE-2023-3439 & 4.7   & v5.17.6 -- v5.17.15 & v5.17.5 & 10 \\
    CVE-2021-33033 & 7.8   & v5.11.7 -- v5.11.13 & v5.11.6 & 7     & CVE-2023-3609 & 7.8   & v6.3.9 -- v6.3.13 & v6.3.8 & 5 \\
    CVE-2021-3411 & 6.7   & v5.9.15 -- v5.9.16 & v5.9.14 & 2     & CVE-2023-3611 & 7.8   & v6.4.5 -- v6.4.16 & v6.4.4 & 12 \\
    CVE-2021-3483 & 7.8   & v5.11.12 -- v5.11.22 & v5.11.11 & 11    & CVE-2023-3812 & 7.8   & v6.0.8 -- v6.0.19 & v6.0.7 & 12 \\
    CVE-2021-3501 & 7.1   & v5.11.16 -- v5.11.22 & v5.11.15 & 7     & CVE-2023-3863 & 4.1   & v6.4.4 -- v6.4.16 & v6.4.3 & 13 \\
    CVE-2021-3573 & 6.4   & v5.12.10 -- v5.12.19 & v5.12.9 & 10    & CVE-2023-4004 & 7.8   & v6.4.7 -- v6.4.16 & v6.4.6 & 10 \\
    CVE-2021-3659 & 5.5   & v5.11.14 -- v5.11.22 & v5.11.13 & 9     & CVE-2023-4128 & 7.8   & v6.4.10 -- v6.4.16 & v6.4.9 & 7 \\
    CVE-2021-3679 & 5.5   & v5.13.6 -- v5.13.19 & v5.13.5 & 14    & CVE-2023-4133 & 5.5   & v6.2.13 -- v6.2.16 & v6.2.12 & 4 \\
    CVE-2021-3732 & 5.5   & v5.13.11 -- v5.13.19 & v5.13.10 & 9     & CVE-2023-4147 & 7.8   & v6.4.8 -- v6.4.16 & v6.4.7 & 9 \\
    CVE-2021-3736 & 5.5   & v5.14.6 -- v5.14.20 & v5.14.5 & 15    & CVE-2023-4389 & 7.1   & v5.17.4 -- v5.18 & v5.17.3 & 13 \\
    CVE-2021-3739 & 7.1   & v5.14.1 -- v5.14.20 & v5.14 & 20    & CVE-2023-4394 & 6.0     & v5.19.6 -- v5.19.17 & v5.19.5 & 12 \\
      \hline
    \end{tabular}%
    }
  \label{tab:appendix-fp-cases}%
\end{table*}%